\documentclass[]{tCPH2e}
\usepackage[colorlinks=true,urlcolor=blue,citecolor=blue,linkcolor=blue,hyperindex]{hyperref}

 \usepackage[usenames]{color}
  \definecolor{dark-green}{RGB}{0,100,0}
  \usepackage{amsmath}
  \usepackage{bm}
  \newcommand{\Espaco}{\rule[-5mm]{0mm}{13mm}} 
  \newcommand{\de}[1]{\left(#1\right)}
  \DeclareMathOperator{\el}{\mbox{\large$e$}}   %EXPONENCIAL LARGE 

\begin{document}
\doi{10.1080/0010751YYxxxxxxxx}
 \issn{1366-5812}
\issnp{0010-7514}

\jvol{00} \jnum{00} \jyear{2010} %\jmonth{June}

\markboth{Taylor \& Francis and I.T. Consultant}{Contemporary Physics}

%\articletype{GUIDE}

\title{{\itshape An introduction to nonadditive entropies and a thermostatistical approach of inanimate and living matter $^\&$ \footnote{$^\&$ Invited review to appear in Contemporary Physics (2014).}}}
%: \LaTeXe\ style guide for authors (`Own Style')}

\author{Constantino Tsallis$^{\ast}$\thanks{$^\ast$
%Corresponding author. 
Email: tsallis@cbpf.br
\vspace{6pt}} \\ \vspace{6pt} {\em{Centro Brasileiro de Pesquisas Fisicas and National Institute of Science and Technology for Complex Systems, Rua Xavier Sigaud 150, 22290-180 Rio de Janeiro-RJ, Brazil}} \\
{\em{and \\ Santa Fe Institute, 1399 Hyde Park Road, Santa Fe, NM 87501, USA} \\
\em{s}}
\\\vspace{6pt}\received{v3.0 released January 2010} 
}

\maketitle

\begin{abstract}
The possible distinction between inanimate and living matter has been of interest to humanity since thousands of years. Clearly, such a rich question can not be answered in a single manner, and a plethora of approaches naturally do exist. However, during the last two decades, a new standpoint, of thermostatistical nature, has emerged. It is related to the proposal of nonadditive entropies in 1988, in order to generalise the celebrated Boltzmann-Gibbs additive functional, basis of standard statistical mechanics. Such entropies have found deep fundamental interest and uncountable applications in natural, artificial and social systems. In some sense, this perspective represents an epistemological paradigm shift. These entropies crucially concern complex systems, in particular those whose microscopic dynamics violate ergodicity. Among those, living matter and other living-like systems play a central role. We briefly review here this approach, and present some of its predictions, verifications and applications.

\bigskip
\bigskip
\bigskip

\hbox to \textwidth{\hsize\textwidth\vbox{\hsize18pc
%\hspace*{-12pt} {1.}    Introduction\\
      }}
\end{abstract}

\section{Introduction}

In 1865, the German physicist and mathematician Rudolf Julius Emanuel Clausius introduced the concept of {\it entropy} (noted $S$, apparently in honour of the French military engineer and physicist Nicolas L\'eonard Sadi Carnot, whom Clausius admired) in order to complete the formalism of thermodynamics. It was defined on macroscopic grounds, in relation to that part of energy that cannot be transformed into work. In the 1870's the Austrian physicist and philosopher Ludwig  Eduard Boltzmann, in a genius insight, connected entropy with the microscopic world. Later on, the American physicist and mathematician Josiah Willard Gibbs refined and extended the connection. The Boltzmann-Gibbs (BG) entropic functional $S_{BG}$ expresses Clausius thermodynamic entropy in terms of the probabilities (noted $\{p_i\}$, and summing up to unity) of occurrence of the possible microscopic configurations compatible with our macroscopic knowledge of the system. In its virtually most simple form, namely corresponding to $W$ discrete possibilities, it is given by
\begin{equation}
S_{BG}=-k \sum_{i=1}^W p_i \ln p_i \;\;\;\Bigl(\sum_{i=1}^W p_i=1\Bigr) \,,
\label{BGentropy}
\end{equation}
$k$ being a conventional constant (usually taken as $k=1$ in information theory, or equal to the Boltzmann constant $k_B$ in thermal physics). In theory of communications, this functional is frequently called Shannon entropy.

This expression becomes, for the particular instance when all probabilities are equal, i.e., $p_i =1/W$, 
\begin{equation}
S_{BG}=k \ln W \,,
\label{BGentropy2}
\end{equation} 
carved on stone on Boltzmann's grave in the Central Cemetery in Vienna. In spite of its formidable simplicity, this expression represents one of the most subtle physical concepts. It is allowed to think that, together with Newton's $F=ma$, Planck's $\epsilon = h \nu$ and Einstein's $E=mc^2$, constitutes an important part of the irreducible core of contemporary physics.

Interestingly enough, the BG expression for entropy plays, as we shall see, a key role in a possible distinction between inanimate and living matter. Also, it is at the heart of what some consider as a {\it paradigm shift}, in the sense of Thomas Kuhn. Indeed, during over one century, that entropy expression has been considered by physicists as universal. For example, we are nowadays accustomed to the fact that energy is {\it not} universal. Indeed, the kinetic energy of a truck is, due to its mass, much larger than that of a fly. Moreover, the Einstein expression for the energy sensibly generalizes the Newtonian one. Nothing like that has emerged in physics during more than one century in what concerns the entropy \footnote{There are of course the important contributions by Claude Shannon in the context of the theory of communications, and by John von Neumann in the context of quantum systems. But none of them modifies the basic {\it logarithmic} nature of the BG measure of lack of information.}. This fact has somehow sedimented in the spirit of many physicists, and other scientists, that Eq. (\ref{BGentropy}) (hence Eq. (\ref{BGentropy2})) is universal, in the sense that it applies to {\it all} systems. {\it We intend to argue here that it is not so}. Indeed, we shall argue that the physical entropic functional connecting the thermodynamic entropy to the microscopic world is {\it not} unique, but depends instead on wide universality classes of (strong) correlations within the elements of the system.

\begin{figure}
\begin{center}
\resizebox*{8cm}{!}{\includegraphics{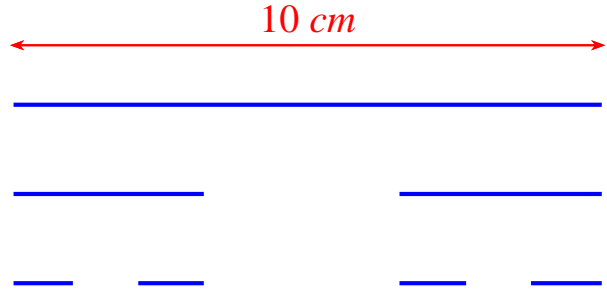}}
%\vspace{0.5cm}
\end{center}
\caption{A triadic Cantor set, whose Hausdorff or fractal dimension is $d_f=\ln 2/ \ln 3=0.6309...$}
\label{cantorset}
\end{figure}

To provide some intuition to the viewpoint that we shall develop here, let us start through a geometrical analogy. Consider the surface of a glass covering a table. Let us consider it, at first approximation, as an Euclidean plane. {\it What is its volume?} Clearly zero since it has no height! A simple answer to a kind of irrelevant question. {\it What is its length?} Clearly infinite, since an infinitely long curve is necessary to entirely cover it! Once again, a simple answer to another kind of irrelevant question. Let us now formulate  a relevant question: {\it What is the area of the surface?} It is say 2 square meters, a {\it finite} number, not zero not infinity. What determines the interesting question --- {\it the area} --- to be addressed in this example? Is it us? Clearly not: it is instead the geometrical nature of our system. As a second illustration along the same lines, let us focus on the triadic Cantor set shown in  Fig. \ref{cantorset}. If we ask say its length, the answer is zero. But if we ask its measure in 0. 6309... dimensions, we obtain a much more informative answer, namely $(10\,cm)^{0.6309...}=4.275...\,cm^{0.6309...}$. As before, it is the specific geometrical structure of the system which determines the useful question to be raised. It is precisely in this sense that the class of dynamical/geometrical correlations between the elements of our system determines the appropriate entropic functional which is to be used. Not that we cannot calculate other functionals, but there is one which is by far the most appropriate and most informative form to use. More specifically, if the system is to be connected with thermodynamics, the appropriate entropic functional is that one which is {\it extensive}. In the next Section we further discuss this point by focusing on the crucial distinction between additivity and extensivity.  Then in Section 3 we discus the time evolution of entropy. In Section 4 we comment extensions of the classical Central Limit Theorem. In Section 5 we present some selected applications, and we finally conclude in Section 6 with some specific epistemological considerations. 

%\begin{figure}
%\begin{center}
%\resizebox*{11cm}{!}{\includegraphics{Thutmosis.eps}}
%\end{center}
%\caption{At the times of Thutmosis I, the North was named, by astronomers and everybody else, ``along the stream", and the South ``against the stream", making of course reference to the highly important river Nile. But great amazement arrived to the Egyptians when, invading Mesopotamia, they found the Eufrates, which runs basically from North to South. Back to Egipt, they registered on an obelisk ``That strange river that, when one goes along the stream, one goes against the stream" (free translation). This extreme confusion was caused by the ignorance that the flow of rivers has nothing to do with the motion of the stars. They were understandably misled by the fact that they had never before seen any other river than the Nile. We intend to metaphorically illustrate here why most physicists have, along almost one century and a half, been confusing entropic additivity and extensivity, misled by the fact that they were basically focusing on systems whose $W$ grows exponentially with $N$. [The images composing this picture are freely available at Internet].}
%\label{Thutmosis}
%\end{figure}

\section{Additive and nonadditive entropic functionals and how they can make the macroscopic entropy to be thermodynamically extensive} 

Let us start by adopting Penrose's definition of entropic {\it additivity} \cite{Penrose1970}. An entropy $S$ is said additive if, {\it for any two probabilistically independent systems} $A$ and $B$ (i.e., such that $p_{ij}^{A+B}=p_i^A p_j^B$, $\forall (i,j)$), we have that  
\begin{equation}
S(A+B)=S(A)+S(B)\,.
\end{equation}
Otherwise it is said nonadditive. It is straightforward to verify from Eq. (\ref{BGentropy}) that $S_{BG}$ is additive.

Let us introduce now two different nonadditive generalisations of $S_{BG}$, namely $S_q$ \cite{Tsallis1988,CuradoTsallis1991,TsallisMendesPlastino1998,Tsallis1999,GellMannTsallis2004,BoonTsallis2005,BeckBenedekRapisardaTsallis2005,AbeHerrmannQuaratiRapisardaTsallis2007,Tsallis2009,Anastasiadis2011,VallianatosTelesca2012} and $S_\delta$ \cite{Tsallis2009,TsallisCirto2013,RibeiroTsallisNobre2013}. 

\subsection{The nonadditive $q$-entropy}
The $q$-entropy is defined through
\begin{equation}
S_q \equiv k\frac{1-\sum_{i=1}^W p_i^q}{q-1} =-k\sum_{i=1}^W p_i^q \ln_q p_i=k\sum_{i=1}^W p_i \ln_q \frac{1}{p_i}  \;\;\;(q \in {\cal R};\, \sum_{i=1}^W p_i=1;\,S_1=S_{BG}) \,,
\label{qentropy}
\end{equation}
where the {\it $q$-logarithmic function} is defined as follows:
\begin{equation}
\ln_q z \equiv \frac{z^{1-q}-1}{1-q} \;\;\;(z>0; \, \ln_1 z= \ln z) \,,
\end{equation}
whose inverse function is defined as follows:
\begin{equation}
e_q^z \equiv [1+(1-q)z]^{\frac{1}{1-q}} \;\;\;(e_1^z= e^z) \,.
\end{equation}
This definition of the {\it $q$-exponential function} applies if $1+(1-q)z \ge 0$; otherwise it vanishes.

The functional form (\ref{qentropy}) has been shown to satisfy uniqueness under appropriate axioms, which generalize, in \cite{Santos1997} and \cite{Abe2000} respectively, those traditionally assumed in the Shannon theorem and in the Khinchin theorem. These interesting generalized theorems, together with the fact that $S_q$ corresponds to the simplest admissible functional which is linear in $\sum_{i=1}^W p_i^q$, and to the facts that it
is both concave and Lesche-stable for $q>0$ (see details in \cite{Tsallis2009}) make this entropy quite interesting a priori.

It can be easily verified that, if $A$ and $B$ are any two probabilistically independent systems, 
\begin{equation}
\frac{S_q(A+B)}{k}= \frac{S_q(A)}{k}+\frac{S_q(B)}{k}+(1-q)\frac{S_q(A)}{k}\frac{S_q(B)}{k} \,,
\end{equation}
hence
\begin{equation}
S_q(A+B)= S_q(A)+S_q(B)+\frac{1-q}{k}S_q(A)S_q(B) \,.
\label{composable}
\end{equation}
We see then that, unless $q=1$, $S_q$ is nonadditive (subadditive for $q>1$, and superadditive for $q<1$). We also see that the $q\to 1$ limit is equivalent to the $k \to\infty$ one. Therefore we expect $q$-statistics to become $q$-independent, hence coincide with BG statistics, in the limit of infinitely high temperatures (we recall that the temperature $T$ always appears in the form $kT$). This property is of the same nature which makes the BG canonical and grand-canonical ensembles to asymptotically coincide with the BG micro-canonical one at high temperatures; and also the Fermi-Dirac and Bose-Einstein statistics to recover, in the same limit, the simple Maxwell-Boltzmann statistics. We also see from Eq. (\ref{composable}) that $S_q$ is {\it composable}, in the sense that, once we know $S_q(A)$, $S_q(B)$ and the index $q$, we do {\it not} need to know microscopic details about the states of $A$ and $B$ to calculate $S_q(A+B)$. Not many entropic functionals have this remarkable property.

If all probabilities are equal we have that
\begin{equation}
S_q=k \ln_q W \,,
\label{qentropy2}
\end{equation}
which generalizes Eq. (\ref{BGentropy2}).

If we are dealing with continuous variables, we have
\begin{equation}
S_q \equiv k\frac{1-\int dx\, [p(x)]^q}{q-1} =-k \int dx\, [p(x)]^q \ln_q [p(x)] =k \int dx\, p(x) \ln_q \Bigl[\frac{1}{p(x)} \Bigr] \;\;\;\Bigl(\int dx\,p(x)=1\Bigr)\,.
\end{equation}
If we are dealing with quantum operators, we have
\begin{equation}
S_q\equiv k\frac{1-Tr [\rho^q]}{q-1}=-kTr [\rho^q \ln_q \rho]=kTr \Bigl[\rho \ln_q \frac{1}{\rho}\Bigr] \;\;\;(Tr \rho=1) \,,
\end{equation}
$\rho$ being the density matrix. The $q=1$ particular case recovers $S_1=-kTr \rho \ln \rho$, currently referred to as the von Neumann entropy.

\subsection{The nonadditive $\delta$-entropy}
The $\delta$-entropy is defined through
\begin{equation}
S_{\delta}\equiv k \sum_{i=1}^W p_i \Bigl[\ln \frac{1}{p_i}\Bigr]^\delta  \;\;\;(\delta \in {\cal R};\, \sum_{i=1}^W p_i=1;\,S_1=S_{BG}) \,.
\label{deltaentropy}
\end{equation}

If all probabilities are equal we have that
\begin{equation}
S_\delta=k [\ln W]^\delta \,,
\label{stretched2}
\end{equation}
which generalizes Eq. (\ref{BGentropy2}), though in a different sense than that of $S_q$. The expressions of $S_\delta$ for the continuous and quantum cases are self-evident.

It can be easily verified that, if $A$ and $B$ are any two probabilistically independent systems, 
\begin{eqnarray}
S_\delta(A+B)&=& k\sum_{i=1}^{W_A} \sum_{j=1}^{W_B} p_i^A p_j^B \Bigl[\ln \frac{1}{p_i^A}+\ln \frac{1}{p_j^B}\Bigr]^\delta \nonumber \\
&\ne& k\sum_{i=1}^{W_A} p_i^A\Bigl[  \ln \frac{1}{p_i^A} \Bigr]^\delta +  k\sum_{j=1}^{W_B} p_j^B\Bigl[  \ln \frac{1}{p_j^B} \Bigr] ^\delta= S_\delta(A)+S_\delta(B) \;\;\;(\delta \ne 1) \,.
\end{eqnarray}
For the particular case of equal probabilities, more precisely $p_i^A=1/W_{A} \,(\forall i)$ and $p_j^B=1/W_{B} (\forall j)$, we obtain
\begin{eqnarray}
S_\delta(A+B)&=&k \Bigl[\ln (W_AW_B) \Bigr]^\delta = \{  [S_\delta(A)]^{1/\delta} +  [S_\delta(B)]^{1/\delta} \}^{\delta} \nonumber \\
&\ne& k \Bigl[\ln W_A \Bigr]^\delta +k \Bigl[\ln W_B \Bigr]^\delta = S_\delta(A) + S_\delta(B) \;\;\;(\delta \ne 1) \,.
\end{eqnarray}
This exhibits that, unless $\delta =1$, $S_\delta$ is nonadditive (subadditive if $\delta < 1$, and superadditive if $\delta >1$).

\subsection{The nonadditive $(q,\delta)$-entropy}
Eqs. (\ref{qentropy}) and (\ref{deltaentropy}) can be unified as follows \cite{TsallisCirto2013,RibeiroNobreTsallis2014}:
\begin{equation}
S_{q,\delta} \equiv k \sum_{i=1}^W p_i \Bigl[\ln_q \frac{1}{p_i} \Bigr]^\delta  \;\;\;(q \in {\cal R},\, \delta \in (\cal R)) \,.
\end{equation}
We straightforwardly verify that $S_{1,1}=S_{BG}$, $S_{q,1}=S_q$, and $S_{1,\delta}=S_\delta$. Clearly, unless $(q,\delta)=(1,1)$, $S_{q,\delta}$ is nonadditive.

There are several other two-parameter entropic functionals available in the literature, e.g., the Borges-Roditi entropy \cite{BorgesRoditi1998}, the Schwammle-Tsallis entropy \cite{SchwammleTsallis2007}, the Hanel-Thurner entropy \cite{HanelThurner2011a,HanelThurner2011b}. Each of them has its own (interesting) motivation. However, the latter deserves a special comment: it has been obtained as the most general trace-form entropy which satisfies the first three axioms of Khinchin \footnote{The four Khinchin axioms essentially are (i) Entropic is continuous in its probability variables; (ii)  Entropy is maximal for equi-distribution; (iii) Adding a zero-probability event does not modify the entropy; (iv) Entropy is additive with regard to the appropriate set of entropy and conditional entropy of two (not necessarily independent) systems $A$ and $B$.}. 

It is given by
\begin{equation}
S_{c,d} \equiv k \Bigl\{\frac{e\sum_{i=1}^W \Gamma(1+d,1-c\ln p_i)}{1-c+cd} -\frac{c}{1-c+cd} \Bigr\} \;\;\;(c \in {\cal R},\, d \in {\cal R}) \,,
\label{cdentropy}
\end{equation}
where $\Gamma$ denotes the incomplete Gamma function. This entropic functional should in principle be, in one way or another (typically asymptotically in some thermodynamical sense), directly related to $S_{q,\delta}$ and to the other two-parameter entropies.  Also, from group-theoretical arguments it has been recently shown \cite{Tempesta2011} that wide classes of entropies are isomorphic to Dirichlet series. In particular, $S_q$ corresponds to the Riemann zeta function. It seems reasonable to expect that all these two-parameter entropies correspond to various Dirichlet series. This remains however to be exhibited explicitly.

\subsection{Entropic extensivity}
{\it Extensivity} is a property completely different from {\it additivity}. We will say that an entropy $S(N)$, $N$ being the number of elements of the system, is extensive if $S(N) \propto N$ for $N\to\infty$, i.e., if $0< \lim_{N\to\infty} \frac{S(N)}{N} < \infty$. Additivity only depends on the mathematical form of the entropic functional, whereas extensivity depends on that {\it and} on the nature of the correlations between the elements of the system. Consequently, it is trivial to check whether a given entropic functional is additive, whereas it can be enormously difficult to check whether a given entropy applied to a given system is extensive.

Let us address now various important classes of systems. We start with those satisfying, in the $N\to\infty$ limit, $W(N) \propto \mu^N  \;(\mu>1)$, i.e., those whose elements are independent or quasi-independent. Eq. (\ref{BGentropy2}) yields $S_{BG}(N) \propto N$, consistently with thermodynamics. In other words, for this class, $S_{BG}$ is extensive and it is therefore the one to be used to study its thermostatistics.

In contrast, strong correlations frequently yield, in the $N\to\infty$ limit, that the number $W$ of microscopic possibilities whose probability is nonzero increases like $W(N) \propto N^\rho \;(\rho>0)$. For such systems, $S_{BG}(N) \propto \ln N$, which is thermodynamically inadmissible. Instead, Eq. (\ref{qentropy2}) implies $S_{q=1-\frac{1}{\rho}}(N) \propto N$, in agreement with classical thermodynamics.

As one more interesting class of correlations, we might have, in the $N\to\infty$ limit, that the number $W$ of microscopic possibilities whose probability is nonzero increases like $W(N) \propto \nu^{N^\gamma} \;(\nu>1, \, 0<\gamma<1)$. For this class, there is no value of $q$ which makes $S_q$ extensive. But, through Eq. (\ref{stretched2}), we have that $S_{\delta=1/\gamma}(N) \propto N$ \cite{TsallisCirto2013}, i.e., it is extensive and it is  therefore the one to be used to address the thermostatistics of the system (for instance, possibly, black holes).

\begin{figure}
\begin{center}
%\begin{minipage}{100mm}
%\subfigure[]{
\resizebox*{11cm}{!}{\includegraphics{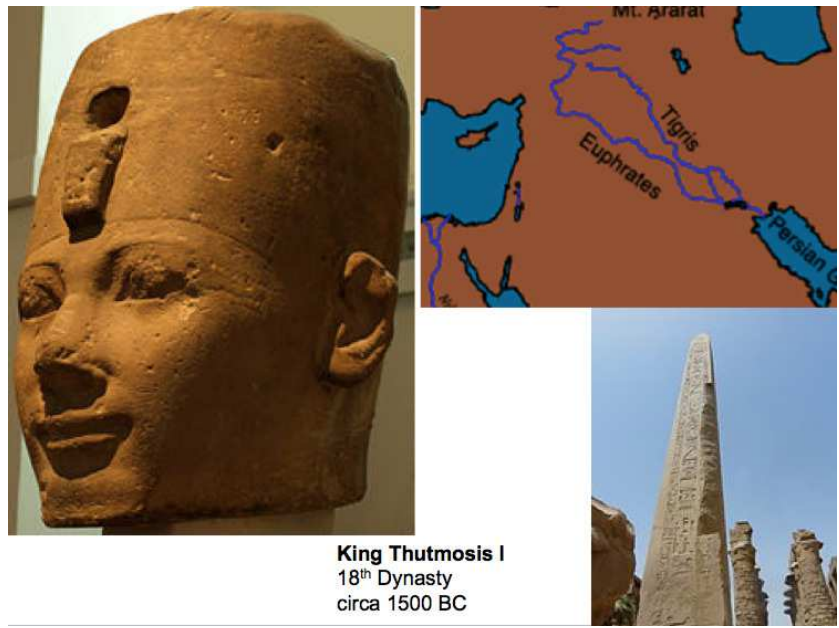}}
%\subfigure[]{
%\resizebox*{5cm}{!}{\includegraphics{senu_gr2.eps}}}%
\end{center}
\caption{At the times of Thutmosis I, the North was named, by astronomers and everybody else, ``along the stream", and the South ``against the stream", making of course reference to the highly important river Nile. But great amazement arrived to the Egyptians when, invading Mesopotamia, they found the Eufrates, which runs basically from North to South. Back to Egipt, they registered on an obelisk ``That strange river that, when one goes along the stream, one goes against the stream" (free translation). This extreme confusion was caused by the ignorance that the flow of rivers has nothing to do with the motion of the stars. They were understandably misled by the fact that they had never before seen any other river than the Nile. We intend to metaphorically illustrate here why most physicists have, along almost one century and a half, been confusing entropic additivity and extensivity, misled by the fact that they were basically focusing on systems whose $W$ grows exponentially with $N$. [The images composing this picture are freely available at Internet].}
%\label{sample-figure}
%\end{minipage}
\label{Thutmosis}
\end{figure}

These considerations are clearly based on the requirement that, {\it in all circumstances}, the thermodynamic entropy should be extensive: see Table  \ref{tableBG} (and also Fig. \ref{Thutmosis}). {\it Why that?} This is a crucially important point by itself, but we shall not address it here in detail because of lack of space. Nevertheless, the main reasons for this (kind of intuitive but by no means trivial) requirement can be seen in \cite{TsallisCirto2013,RuizTsallis2013}. In \cite{TsallisCirto2013} we summarize a strong thermodynamical reason, based essentially in the Legendre transformations that characterize thermodynamics. In \cite{RuizTsallis2013} we illustrate, with a specific model, the large-deviation-theory reason. Both reasons are consistent among them, as their detailed analysis can show. For some details concerning the Legendre-transform reason the reader might look at \cite{Tsallis2009}.  

We notice that, in the limit $N\to\infty$, $\mu^N \gg \nu^{N^\gamma} \gg N^\rho$, which essentially means that the occupancy of the exponential class yields a {\it finite Lebesgue measure}, typical of {\it ergodicity} in full phase space (as imagined by Boltzmann himself consistently with his {\it molecular chaos hypothesis}, as well as by Gibbs), whereas the occupancy of both the stretched-exponential and the power-law classes yield a {\it zero Lebesgue measure}, typical of systems which are {\it nonergodic} in full phase space. It is through this viewpoint that a crucial distinction emerges between {\it inanimate matter} (typically corresponding to the exponential class) and complex systems (typically corresponding to classes such as the stretched-exponential and power-law ones), within which we find {\it living matter}.

\begin{table}
\centering
\begin{tabular}{|c|c|c|c|}   \cline{2-4} 
%-----------------------------------------------------------------------
\multicolumn{1}{c|}{}                                &\multicolumn{3}{c|}{\rule[-3mm]{0mm}{9mm}\footnotesize{\textbf{ENTROPY}} } \\  \hline
\rule[-1mm]{0mm}{7mm}$\bm{W\de{N}}$                  &                   $S_{BG}$                                 & $S_{q}$                                                       & $S_{\delta}$                                                 \\
\rule[-2mm]{0mm}{7mm}$\bm{\de{N \to \infty}}$        &                                                            & $\de{q\neq1}$                                                 & $\de{\delta\neq1}$                                           \\
\rule[-3mm]{0mm}{7mm}                                & \textcolor{dark-green}{\textbf{\footnotesize{(ADDITIVE)}}} & \textbf{\textcolor{dark-green}{\footnotesize{(NONADDITIVE)}}} &\textbf{\textcolor{dark-green}{\footnotesize{(NONADDITIVE)}}} \\ \hline 
%-----------------------------------------------------------------------
\Espaco$\displaystyle{{\sim\mu^{N}} \atop \de{\mu > 1} }$  &\textbf{\textcolor{blue}{\footnotesize{EXTENSIVE}}}   & \textbf{\textcolor{red}{\footnotesize{NONEXTENSIVE}}}  &\textbf{\textcolor{red}{\footnotesize{NONEXTENSIVE}}}              \\ \hline
\Espaco$\displaystyle{\sim N^{\rho} \atop \de{\rho > 0} }$ &\textbf{\textcolor{red}{\footnotesize{NONEXTENSIVE}}} & \textbf{\textcolor{blue}{\footnotesize{EXTENSIVE}}}    &\textbf{\textcolor{red}{\footnotesize{NONEXTENSIVE}}}              \\
\rule[-2mm]{0mm}{0mm}                                      &                                                      & $\de{q = 1- 1/\rho}$                                   &                                                                   \\ \hline
\Espaco$\displaystyle{\sim\nu^{ N^\gamma } \atop (\nu>1; }$&\textbf{\textcolor{red}{\footnotesize{NONEXTENSIVE}}} & \textbf{\textcolor{red}{\footnotesize{NONEXTENSIVE}}}  & \textbf{\textcolor{blue}{\footnotesize{EXTENSIVE}}}               \\
\rule[-2mm]{0mm}{0mm}$0<\gamma<1)$                         &                                                      &                                                        & $\de{\delta = 1/\gamma}$                                          \\ \hline
%-----------------------------------------------------------------------
\end{tabular}
\vspace{0.7cm}
\caption{Additive and nonadditive entropic functionals and classes of systems for which the entropy is extensive. $W(N)$ is the number of admissible microscopic configurations of a system with $N$ elements; only configurations with nonvanishing occurrence probability are considered admissible.}
\label{tableBG}
\end{table}

\subsection{Calculation of $q$ from first principles}

\begin{figure}
\begin{center}
%\begin{minipage}{100mm}
%\subfigure[]{
\resizebox*{12cm}{!}{\includegraphics{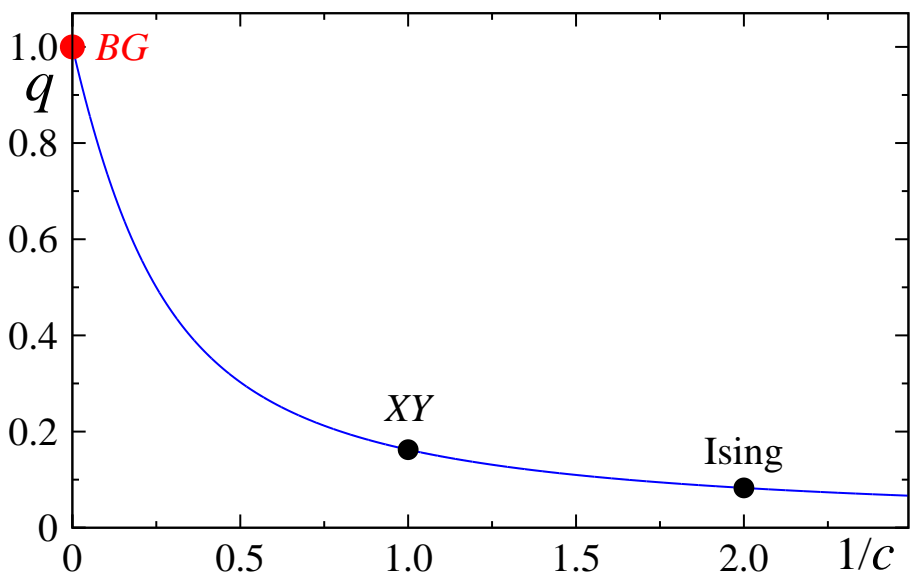}}
%\subfigure[]{
%\resizebox*{5cm}{!}{\includegraphics{senu_gr2.eps}}}%
\end{center}
\caption{The block BG entropy $S_{BG}(L) \propto \ln L$ for all finite values of the central charge $c$, thus violating thermodynamical extensivity, whereas, for the special value of $q$ indicated here,  $S_q(L) \propto L$, i. e., it satisfies one-dimensional extensivity,  thus enabling the use of all the relations present in any good textbook of thermodynamics. See details in \cite{CarusoTsallis2008}.}
%\label{sample-figure}
%\end{minipage}
\label{centralcharge}
\end{figure}

The entropic index $q$ is to be calculated from the geometric/dynamical nature of the occupancy of phase space. For example, if we have a many-body Hamiltonian system, $q$ is determined by the Hamiltonian itself (more precisely, by the universality class of the Hamiltonian), i.e., from first principles. This calculation is by no means a mathematically easy task, and in many cases it is practically intractable. A few examples have, nevertheless, been analytically worked out. Such are the systems addressed in \cite{CarusoTsallis2008}. Let us consider a strongly quantum-entangled system constituted by $N$ elements with no frustrated interactions. Let us further assume that, at the $N\to\infty$ limit, it exhibits a $T=0$ critical point, i. e., that it undergoes a quantum phase transition. Being at $T=0$, its quantum state is a {\it pure} one, hence its entropy -- calculated through any reasonable entropic functional --- vanishes. If we denote by $\rho_N$ its density matrix, we have that $Tr \rho_N^2 =Tr \rho_N=1$. If we select now $L$ of its $N$ elements and trace out $(N-L)$ elements, i. e., if we define $\rho_{L,N} \equiv Tr_{N-L} \rho_N$, we will generically have a {\it mixed} state because $Tr \rho_{L,N}^2 < Tr \rho_{L,N}=1$. We have then here, due to quantum nonlocality, the very curious, nonclassical, situation where the entropy, say the $q$-entropy for any fixed value of $q$, of a subsystem, called sometimes {\it block entropy}, is positive, hence {\it larger} than the entropy of the entire system, which vanishes!  To be more precise, in such situations we have that the limiting entropy $\lim_{N\to\infty} S_q[\rho_N]$ vanishes, whereas $S_q(L) \equiv \lim_{N\to\infty} S_q[\rho_{L,N}]$ is positive and depends on $L$.  It would of course be extremely convenient if a special value of $q$ exists such that $S_q(L) \propto L$. Indeed, satisfaction of the thermodynamical extensivity requirement will allow us to freely use all the relations that appear in any good textbook of thermodynamics! The peculiar situation that we have described here indeed occurs for the following Hamiltonian of spins-1/2 interacting ferromagnetically between first-neighbors along a linear chain:
\begin{equation}
{\cal H}=- \sum_{j=1}^{N-1} \Bigl[(1+\gamma) \sigma_j^x \sigma_{j+1}^x + (1-\gamma)   \sigma_j^y \sigma_{j+1}^y \Bigr] -2\lambda \sum_{j=1}^N\sigma_j^z \,,
\end{equation}
$\lambda$ being a transverse field, and the $\sigma$'s being Pauli matrices; $|\gamma|=1$ corresponds to the Ising ferromagnet (whose central charge $c$ equals 1/2 
\footnote{The central charge $c$ is a nonnegative real number which plays an important role in Virasoro algebra and conformal quantum field theory by determining the form of the correlation functions. In the interval $(0,1)$ $c$ takes discrete values which accumulate at $c=1$, and which characterize critical universality classes for two-dimensional classical models; above $c=1$, it varies continuously up to infinity, where quantum effects (hence quantum nonlocality) disappear.}
, as well known), $0<|\gamma|<1$ corresponds to the anisotropic XY ferromagnet, belonging to the same universality class of the Ising model, and $\gamma =0$ corresponds to the isotropic XY ferromagnet, a different universality class (whose central charge $c$ equals unity, as well known).  The particular case that is being focused is the $(T, \lambda)=(0,1)$ quantum critical point. By extending the arguments to the generic two-dimensional universality classes characterized by a central charge $c \ge 0$, it has been analytically established that the special value of $q$ which produces an {\it extensive} {\it nonadditive} block entropy $S_q(L)$ is given by \cite{CarusoTsallis2008}:
\begin{equation}
q= \frac{\sqrt{9+c^2} -3}{c}\,.
\label{analyticq}
\end{equation}
This entropic index monotonically increases from zero to the BG value $q=1$ when $c$ increases form zero to infinity: see Fig. \ref{centralcharge}. 
Another example (though only numerical) of $q(c)$ producing an extensive $S_q(L)$ can be seen, for a different magnetic system, in \cite{SaguiaSarandy2010}.

It is at this point worthy to mention that here and there $q$-statistics is occasionally criticized by stating that the index $q$ is nothing but a fitting parameter with no physical or mathematical meaning whatsoever. As we have above illustrated, such statements are the mere consequence of pure ignorance. The situation might be seen as analogous to Newtonian mechanics. Indeed the actual orbit of Mars has not been analytically obtained from first principles, {\it not} because classical mechanics is an incomplete theory, but just because of the impossibility of having all the initial positions and velocities of all masses of the planetary system (and even if we did have them, we would not have access to a computer powerful enough to handle all this information in a gigantic set of nonlinear Newtonian equations). What astronomers then do is to use the correct (at first approximation) Keplerian {\it mathematical form} (namely an ellipse), that has been deduced within Newtonian mechanics, and then establish the rest of the parameters through fitting good-quality astronomical data. In the handling of actual data of many complex systems, it is appropriate to use the $q$-exponential {\it mathematical form}, $q$ being established through fitting just because either the precise dynamics of the system is unknown, or, even when it is known, it is mathematically intractable, {\it not} because $q$-statistics is an incomplete theory (it is thinkable that, like any other theory, it could be wrong, but definitively not incomplete!).

\section{Time evolution of the entropy}

\begin{figure}
\begin{center}
%\begin{minipage}{100mm}
%\subfigure[]{
\resizebox*{7.0cm}{!}{\includegraphics{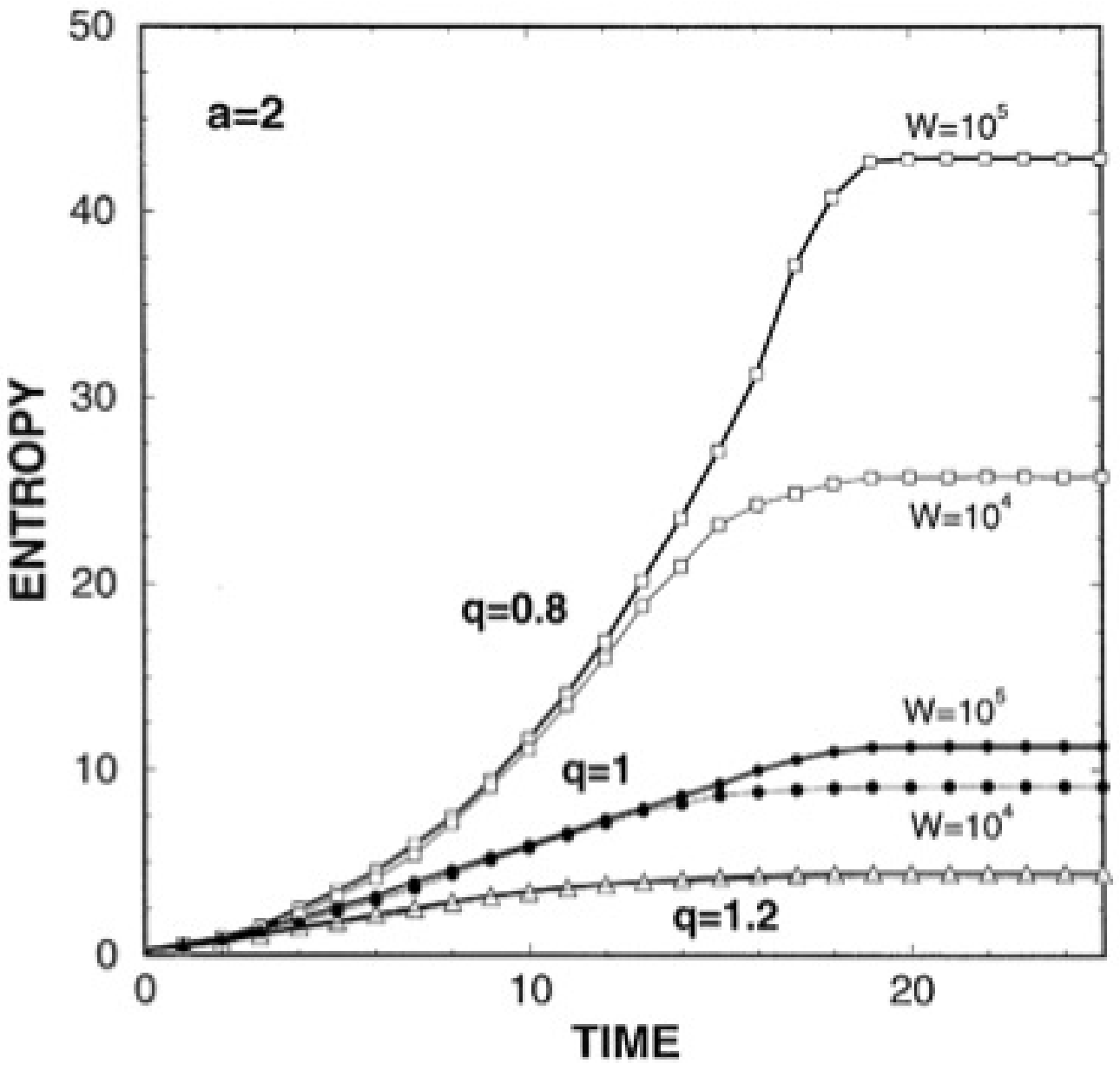}}
\resizebox*{7.3cm}{!}{\includegraphics{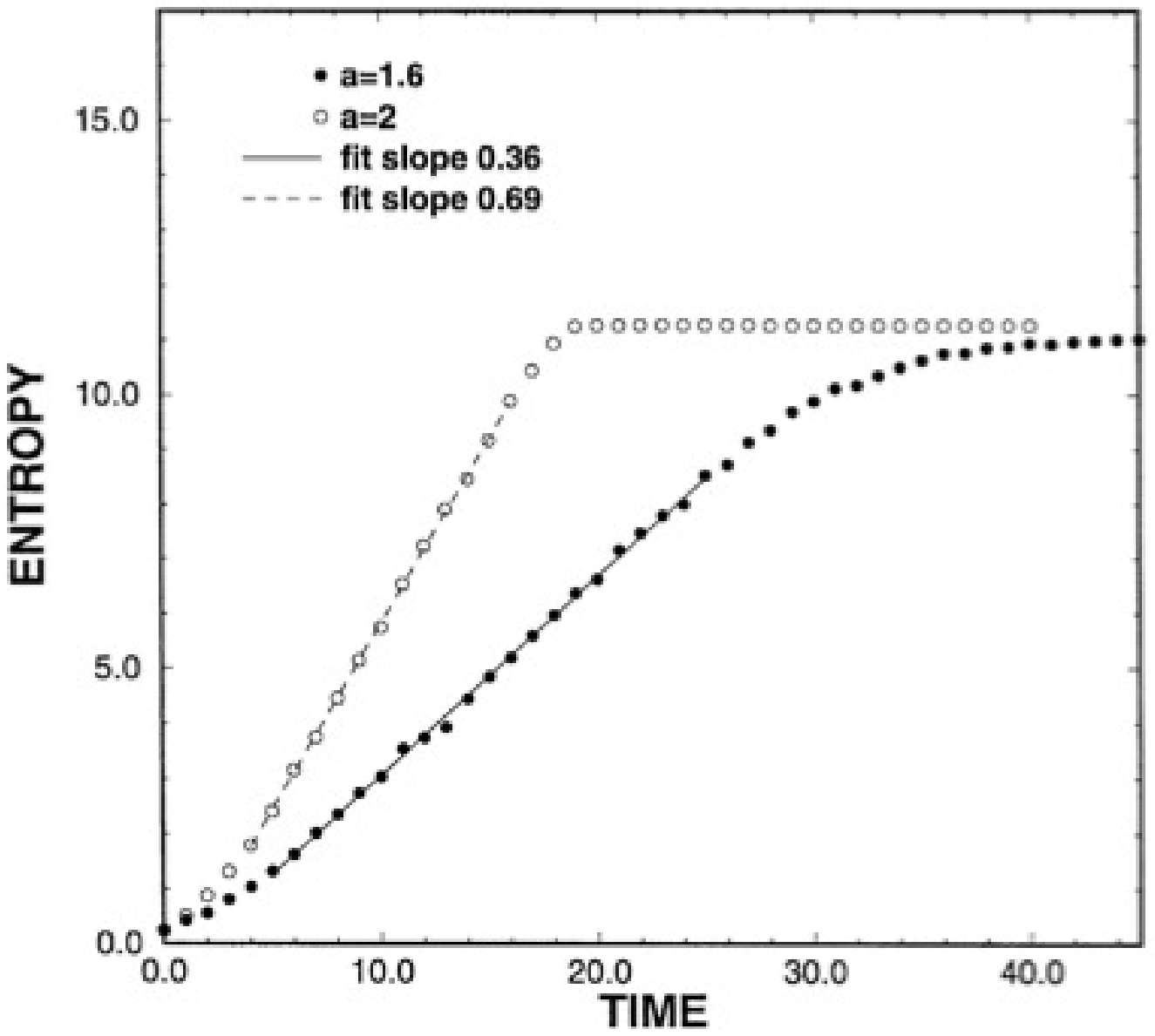}}
%\subfigure[]{
%\resizebox*{5cm}{!}{\includegraphics{senu_gr2.eps}}}%
\end{center}
\caption{The time-dependence of the entropy $S_q$ for typical values of the index $q$ for $a=2$ and $a=1.6$. The relevant (finite) slope is in this case given by $K_1 \equiv \lim_{t\to\infty} \lim_{W\to\infty} \lim_{N\to\infty} \frac{S_{BG}(t)}{t}$, where $N$ is the number of initial conditions $\{ x_0\}$ that have been used in order to perform the numerical computation, and $W$ is the number of windows within which the $N$ values of $x_t$ evolve in the interval $[-1,1]$. The Lyapunov exponent for $a=2$ equals $\ln 2 \simeq 0.69.$
See details in \cite{LatoraBarangerRapisardaTsallis2000}.}
%\label{sample-figure}
%\end{minipage}
\label{tqentropy}
\end{figure}

The entropy of a given dynamical system generically depends on both the number $N$ of its elements and time $t$. In the previous Section we have discussed on how it depends on $N$. Let us now discuss on how it depends on $t$. We will see that the $t$-dependence is totally analogous to the $N$-dependence, $t$ and $N$ playing a strongly analogous role (see also \cite{TsallisGellMannSato2005}).
We shall here illustrate the main concepts on a simple example, the well known logistic map:
\begin{equation}
x_{t+1}=1-a x_t^2 \;\;\;(t=0,1,2,...;\, -1 \le x_t \le 1; \, 0 \le a \le 2).
\end{equation}
This is a one-dimensional dissipative map which, depending on the value of the control parameter $a$, the (unique) {\it Lyapunov exponent} $\lambda_1$ can be positive ({\it strong chaos}), zero or negative (the meaning of the subindex $1$ will soon become transparent), $\lambda_1$ being defined a few lines here below. For increasing $a<a_\infty=1.401155189...$, the attractors undergo successive bifurcations which accumulate when approaching the value $a_\infty$. At $a=a_\infty$, $\lambda_1=0$ ({\it weak chaos}); at $a=2$, the system achieves its most chaotic point (i.e., its largest possible value for $\lambda_1$), for which $\lambda_1=\ln 2 = 0.6931...$. For this value of $a$ and for a smaller one, the time-dependence of the entropy $S_q$ is as shown in Fig. \ref{tqentropy} (from \cite{LatoraBarangerRapisardaTsallis2000}). 
The $W\to\infty$ limit of $S_q(t)$ is a monotonically increasing function whose $t\to\infty$ slope diverges for $q<1$, vanishes for $q>1$, and is {\it finite} for $q=1$, i.e., for $S_{BG}$. It is in this manner that the system itself determines what entropic functional to use in order to most efficiently quantify its degree of disorder or uncertainty. The finite slope is referred to as the {\it entropy production per unit time} and is noted $K_1$ (this concept quite generically coincides with the so called Kolmogorov-Sinai entropy rate). 

Let us now define precisely the Lyapunov exponent. The divergence of two infinitely close initial conditions is characterized by the {\it sensitivity to the initial conditions} $\xi(t) \equiv \lim_{\Delta x(0) \to 0} \frac{\Delta x(t)}{\Delta x(0)}$. Whenever the system is strongly chaotic we verify
\begin{equation}
\xi = e^{\lambda_1\,t}\,,
\end{equation}
where $\lambda_1 >0$ is the Lyapunov exponent. We verify in the logistic map case a Pesin-like identity, namely
\begin{equation}
K_1 =\lambda_1 \,,
\end{equation}
if $\lambda_1 >0$, and zero otherwise (more precisely if $\lambda_1 \le 0$). Since, for $N \gg W \gg t \gg 1$, we have that $S_{BG} \propto t$, we may measure time in terms of bits. This provides a novel manner for thinking the concept of {\it time} and the possible time reversibility (or irreversibility) of the ``most microscopic law of nature" \footnote{This and related issues have been the object of historical philosophical debates between A. Einstein and H. Bergson, among many others. For example, I. Prigogine states that ``Nature's real equation must not be invariant with regard to time inversion" (see his articles with I. Stengers, as well as his book {\it From Being to Becoming}). Einstein, for example, states the opposite. The question has been focused on by many others, such as P. Langevin, H. Poincar\'e, G. Deleuze, %(Bergsonisme)
F. Guattari, 
%(La Revolution Moleculaire, Chaosmosis)
B. Latour.}. 

\begin{figure}
\begin{center}
%\begin{minipage}{100mm}
%\subfigure[]{
\resizebox*{12cm}{!}{\includegraphics{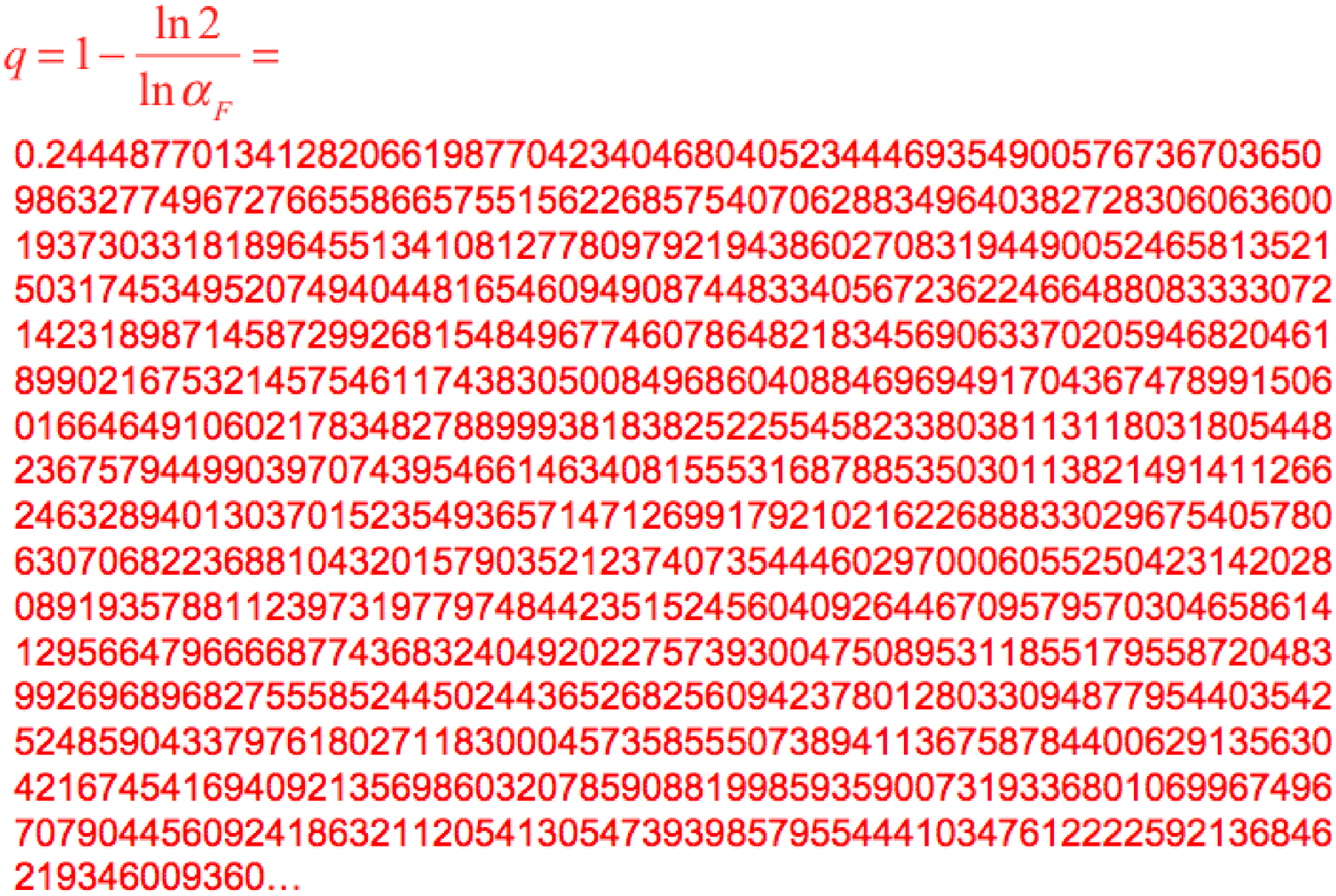}}
%\subfigure[]{
%\resizebox*{5cm}{!}{\includegraphics{senu_gr2.eps}}}%
\end{center}
\caption{The first 1018 digits of the value of the entropic index $q$ which characterizes the sensitivity to the initial conditions and the entropy production per unit time of the logistic-map at the Feigenbaum point $a_\infty$ (accumulation point of the successive bifurcations). It has been calculated, through Eq. (\ref{qalpha}), by using the 1018 digits of the Feigenbaum universal constant $\alpha_F$ freely available at  
\url{http://pi.lacim.uqam.ca/piDATA/feigenbaum.txt} 
}
%\label{sample-figure}
%\end{minipage}
%\end{center}
\label{qvalue}
\end{figure}

\begin{figure}
\begin{center}
%\begin{minipage}{100mm}
%\subfigure[]{
\resizebox*{11cm}{!}{\includegraphics{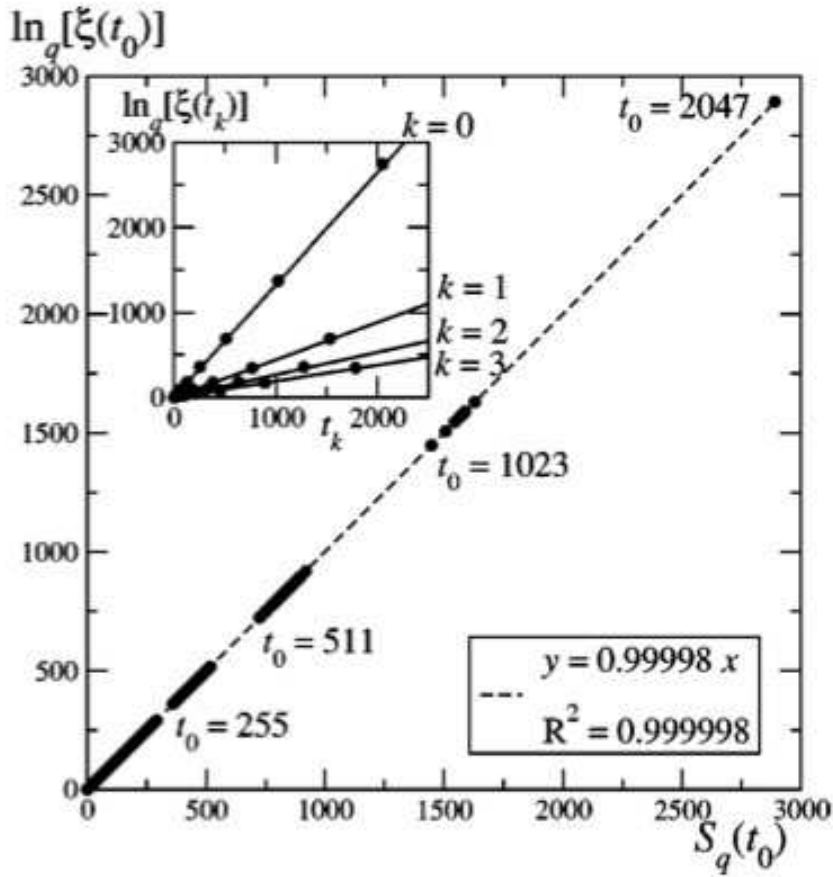}}
%\subfigure[]{
%\resizebox*{5cm}{!}{\includegraphics{senu_gr2.eps}}}%
\end{center}
\caption{Time-dependence, at the Feigenbaum point (weak chaos), of the entropy $S_q(t)$ and of the sensitivity to the initial conditions $\xi(t)$, with $q=0.24448...$ given by Eq. (\ref{qalpha}). For details see \cite{BaldovinRobledo2004}.
}
%\label{sample-figure}
%\end{minipage}
\label{tqentropy3}
\end{figure}

Instead of focusing on $a=2$ and other values of $a$ for which $\lambda_1 >0$, let us focus now on $a=a_\infty$ (the so called Feigenbaum point), for which $\lambda_1$ vanishes, hence $K_1=0$. We verify that \cite{TsallisPlastinoZheng1997,BaldovinRobledo2004}
\begin{equation}
\xi = e_q^{\lambda_q^{k}\,t_k}\,,
\end{equation}
where $q$ is a special value (given in Fig. \ref{qvalue}), and the set $\{\lambda_q^{k}\}$ is related to $1/(1-q)$ as indicated in Fig. \ref{tqentropy3} \cite{BaldovinRobledo2004}. We also verify that, in the asymptotic $t\to\infty$ limit, $S_q(t)$ has zero (infinite) slope for all values of $q$ different from that one, whereas, for that value of $q$, it has {\it finite} slopes $K_q^{k}$ satisfying precisely the following $q$-generalized Pesin-like identity:
\begin{equation}
K_q^{k}=\lambda_q^{k} \,.
\end{equation}
This unique value of $q$ has been shown \cite{LyraTsallis1998} to satisfy
\begin{equation}
\frac{1}{1-q}=\frac{1}{\alpha_{min}} - \frac{1}{\alpha_{max}} \,,
\end{equation}
where $\alpha_{min}$ and $\alpha_{max}$ are respectively the small and large values of $\alpha$ at which the multifractal function $f(\alpha)$ vanishes. From this relation it straightforwardly follows  that
\begin{equation}
q= 1-\frac{\ln 2}{\ln \alpha_F} \,,
\label{qalpha}
\end{equation}
where $\alpha_F$ is the well known Feigenbaum universal constant.

As we see, it is only $S_q$ for this special value of $q$ that satisfies $S_q \propto t$, and which therefore enables (similarly to what happens with $S_{BG}$  for those values of $a$ whose Lyapunov exponent is {\it positive}) to measure time in terms of bits. The entropy $S_{BG}(t)$, although computable, is, in this sense, useless. 

The present value of $q$ in Fig. \ref{qvalue} and that of Eq. (\ref{analyticq}) constitute two paradigmatic examples of analytical calculation of the $q$-index from first principles, the first one corresponding to a low-dimensional-phase-space dissipative (classical) system, and the second one to a high-dimensional-phase-space conservative (quantum) system.

\section{Central Limit Theorem and related questions}

\subsection{q-generalized Central Limit Theorems (CLT) and Fokker-Planck equations (FPE)}

It is interesting to consider the following quite general Fokker-Planck-like equation:
\begin{equation}
\frac{\partial^{{\bar \beta}} p(x,t)}{\partial t^{\bar \beta}} = D_{q,\alpha} \frac{\partial^\alpha [p(x,t)]^{2-q}}{\partial |x|^\alpha} \;\;\;(t \ge 0; \,0< {\bar \beta} \le 1;\, 0<\alpha \le 2; \, q<3) \,,
\label{FP}
\end{equation}
$D_{q,\alpha}$ being a generalized diffusion coefficient. This coefficient will, from now on, be incorporated into time $t$; consequently, without loss of generality we can consider $|D_{q,\alpha}|=1\;(q \ne 2)$. We have therefore three relevant variables, namely  $(q,\alpha,{\bar \beta})$. For our present purposes, we shall only consider the case ${\bar \beta}=1$, and $p(x,0)=\delta(x)$, where $\delta(x)$ denotes the Dirac delta.

\begin{figure}
\begin{center}
%\begin{minipage}{100mm}
%\subfigure[]{
\resizebox*{14cm}{!}{\includegraphics{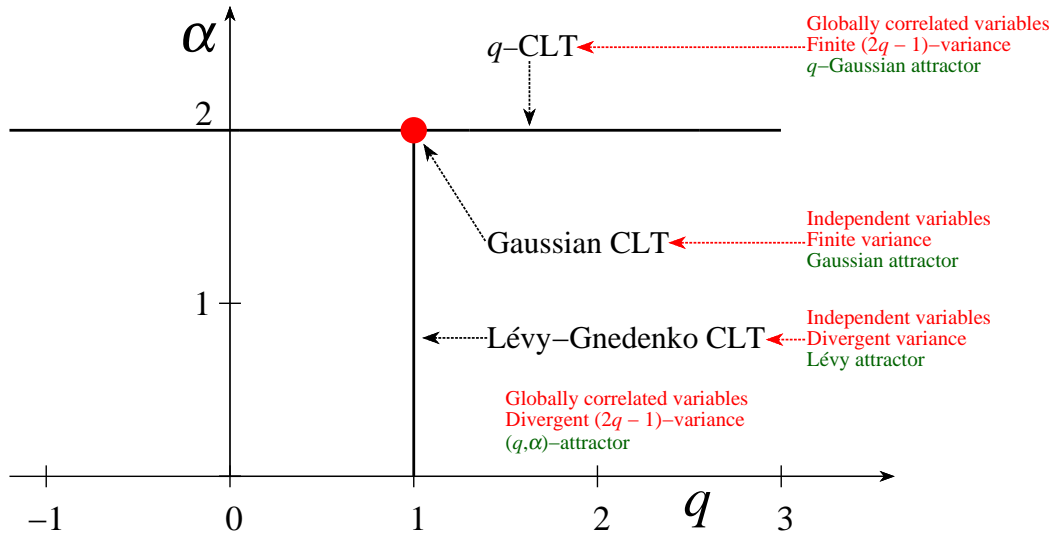}}
%\subfigure[]{
%\resizebox*{5cm}{!}{\includegraphics{senu_gr2.eps}}}%
\end{center}
\caption{Diagram for the Eq. (\ref{FP}) for ${\bar \beta}=1$. It was conjectured in \cite{BolognaTsallisGrigolini2000,Tsallis2005} that the various types of solutions of the FPE would precisely correspond to the attractors (in probability space) of the CLT and its extensions (in red and green). The $(q,\alpha)$-attractor corresponds to {\it globally correlated variables} and {\it divergent $q$-variance}, whose precise definitions are given in \cite{UmarovTsallisSteinberg2008,UmarovTsallisGellMannSteinberg2010}. 
The conjecture of the correspondence between the solutions of the FPE and the attractors of the CLT has proved valid in all the cases that have been checked: it remains to be checked for the generic $(q,\alpha)$-attractors (this is under progress now).
}
%\label{sample-figure}
%\end{minipage}
\label{FPE}
\end{figure}

If $(q,\alpha)=(1,2)$ we have Fourier's well known heat transfer equation, whose solution is $p(x,t) \propto e^{- x^2/(2t)}$. If $q=1$ and $\alpha<2$ we have $p(x,t)$ given by L\'evy's distributions, which asymptotically decay as $1/|x|^{\alpha + 1}$. If $\alpha=2$ and $1<q<3$ we have \cite{PlastinoPlastino1995,TsallisBukman1996} $p(x,t) \propto e_q^{-\beta_q x^2/t^{2/(3-q)}}$ with $\beta_q >0\;(\beta_1=1/2)$, which asymptotically decay as $1/|x|^{2/(q-1)}$. For the generic case $q\ge 1$ and $\alpha \le 2$, we expect the $(q,\alpha)$-stable distributions (to be mentioned later on) which recover all the previous distributions as particular instances, in particular L\'evy distributions for $q=1$ and $\alpha <2$, and $q$-Gaussians for $\alpha=2$ and $q \ge 1$. See Fig. \ref{FPE}.
It is important to realize that there is an important connection between Fokker-Planck-like equations and entropic forms. More specifically, if we impose the validity of the H-theorem (directly related to the validity of second principle of thermodynamics), we obtain a strict relation between the functions that are present in the partial derivative equation and the functional form of the entropy (details can be seen in 
\cite{Frank2005,
%FrankDaffertshofer1999,
Shiino2001,NobreCuradoRowlands2004,
%FrankDaffertshofer2001,
%Chavanis2003,
Chavanis2008,
%SchwammleNobreCurado2007,
%SchwammleCuradoNobre2007,
SchwammleCuradoNobre2009,
RibeiroNobreCurado2011,
%CuradoNobre2003,
RibeiroTsallisNobre2013,RibeiroNobreTsallis2014}). 
One of the remarkable --- and very desirable --- consequences of this intimate relation is the fact that the stationary state of the Fokker-Planck-like equation in the presence of any confining potential  {\it precisely coincides} with the distribution which is obtained by extremizing, under appropriate constraints, the corresponding entropy.

The attractors in the same sense (i.e., after centering and rescaling of sums of an infinitely large number $N$ of random variables) of the classical CLT have been proved to follow the same scenario (see Table \ref{CLT}) as the solutions of the FPE.
\begin{table}
\begin{center}
%\begin{minipage}{100mm}
%\subfigure[]{
\resizebox*{16cm}{!}{\includegraphics{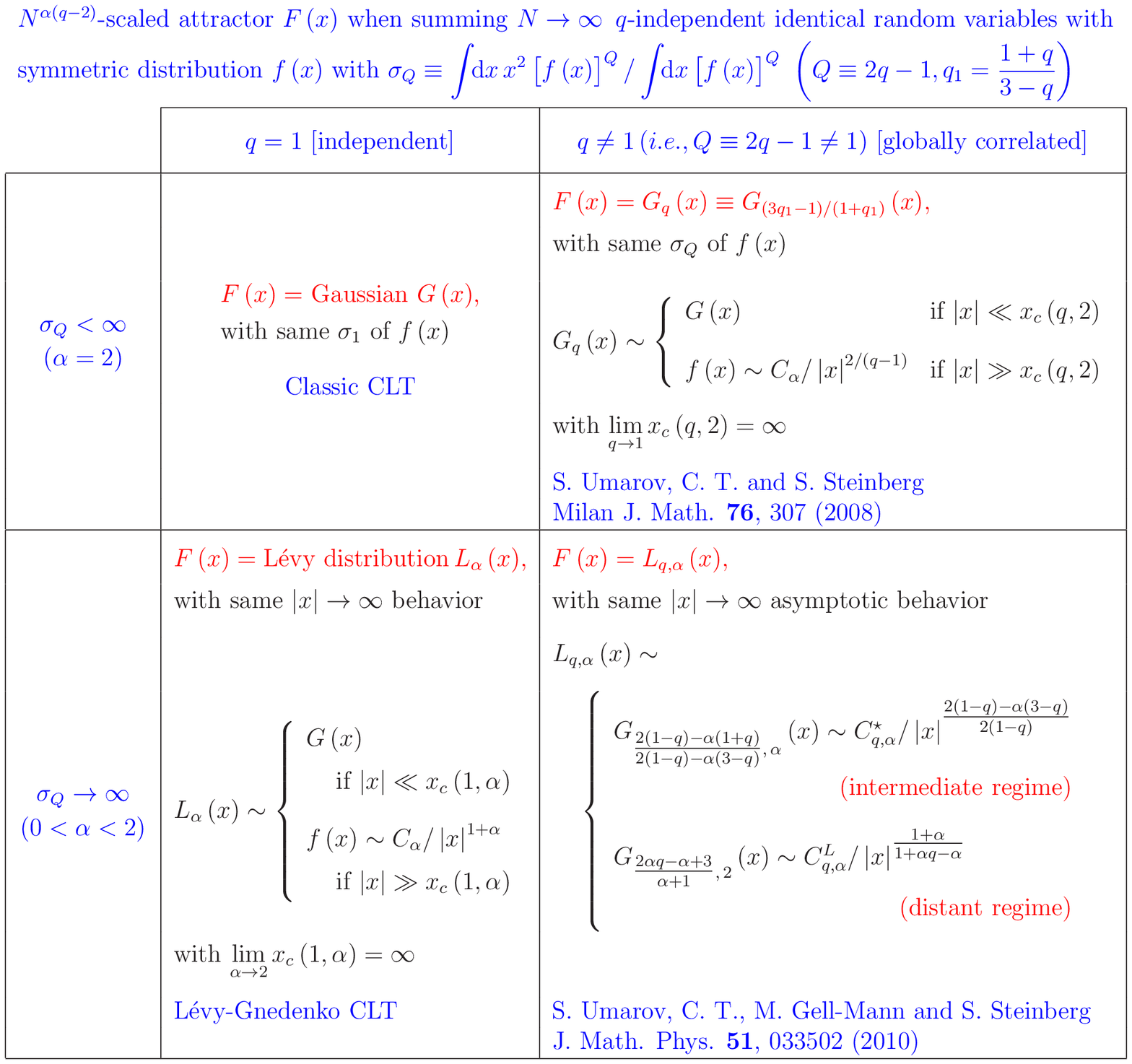}}
%\subfigure[]{
%\resizebox*{5cm}{!}{\includegraphics{senu_gr2.eps}}}%
\end{center}
\caption{Schematic synopsis of the various extensions of the classical CLT. See details in \cite{UmarovTsallisSteinberg2008,UmarovTsallisGellMannSteinberg2010}. Notice that, excepting the Gaussians, all cases asymptotically yield power-laws, the L\'evy distributions being only one of them. There are in fact infinitely many forms of probability distributions that asymptotically behave as power-laws. It is therefore a strong, though widely spread in the literature, confusion to refer to all of them as {\it L\'evy distributions}.
}
%\label{sample-figure}
%\end{minipage}
\label{CLT}
\end{table}

%\subsection{Large deviation theory (LDT)}

\subsection{$q$-triplet}

Let us briefly address here a concept --- the {\it $q$-triplet} or {\it $q$-triangle} --- that frequently emerges in the analysis of complex systems. Consider the following differential equation:
\begin{equation}
\frac{dy}{dx}=ay\;\;\;(y(0)=1)\,,
\label{BGdiff}
\end{equation}
the solution being $y=e^{a\,x}$. This equation can be generalized as follows:
\begin{equation}
\frac{dy}{dx}=ay^q\;\;\;(y(0)=1)\,,
\label{qdiff}
\end{equation}
the solution being $y=e_q^{a\,x}$. These simple facts can be given at least three different physical interpretations, namely related to sensitivity to the initial conditions (characterized by $q_{sen}$), distribution of the stationary state (characterized by $q_{stat}$), and relaxation (characterized by $q_{rel}$) \cite{Tsallis2004}. Within BG statistical mechanics we typically have $q_{sen}=q_{stat}=q_{rel}=1$. See Table \ref{qtriplet}.

\begin{table}
\centering
\begin{tabular}{|p{3.7cm}|c|c|l|} \cline{2-4}
% \begin{tabular}{|l|c|c|c|} \cline{2-4}
%-----------------------------------------------------------------------
\multicolumn{1}{c|}{\rule[-3mm]{0mm}{9mm}} &  $x$ & $a$  & $y\de{x}$ \\ \hline
%-----------------------------------------------------------------------
\multicolumn{4}{|c|}{\rule[-4mm]{0mm}{11mm} \textbf{Boltzmann-Gibbs statistical mechanics ($\bm{S_{BG}}$)}} \\ \hline
%-------------------
% \Espaco \footnotesize{\textbf{Equilibrium distribution}}                                                                                          & $E_i$ & $-\beta$ & $Z p\de{E_i} = $  \\\hline
  \Espaco \raisebox{1.5ex}{\footnotesize{\textbf{Thermal equilibrium}}} \raisebox{-1.5ex}{\hspace{-3.24cm}\footnotesize{\textbf{distribution}}}             & $E_i$ & $-\beta$ & $Z p\de{E_i} = \el^{\displaystyle{-\beta E_i}}$  \\\hline
% \Espaco \footnotesize{\textbf{Sensitivity to initial conditions}}                                                                                 & $t$ & $\lambda$ & $\xi\equiv\dfrac{\Delta x\de{t}}{\Delta x\de{0}} = \el^{\displaystyle{\lambda t}}$  \\\hline
  \Espaco \raisebox{1.5ex}{\footnotesize{\textbf{Sensitivity to initial}}} \raisebox{-1.5ex}{\hspace{-3.13cm}\footnotesize{\textbf{conditions}}}    & $t$ & $\lambda$ & $\xi\de{t}\equiv\dfrac{\Delta x\de{t}}{\Delta x\de{0}} = \el^{\displaystyle{\lambda t}}$  \\\hline
% \Espaco \footnotesize{\textbf{Typical relaxation of observable $O$} }                                                                             & $t$ & $-1/\tau$ & $\Omega\de{t} = \dfrac{O\de{t} - O\de{\infty}}{O\de{0} - O\de{\infty}} = \el^{\displaystyle{-t/\tau}}$  \\\aline
  \Espaco \raisebox{1.5ex}{\footnotesize{\textbf{Typical relaxation}}} \raisebox{-1.5ex}{\hspace{-2.84cm}\footnotesize{\textbf{of observable $O$}}}  & $t$ & $-1/\tau$ & $\Omega\de{t} = \dfrac{O\de{t} - O\de{\infty}}{O\de{0} - O\de{\infty}} = \el^{\displaystyle{-t/\tau}}$  \\\hline
%-----------------------------------------------------------------------
\multicolumn{4}{|c|}{\rule[-4mm]{0mm}{11mm} \textbf{Nonextensive statistical mechanics ($\bm{S_{q}}$)}} \\ \hline
%-------------------
% \Espaco \footnotesize{\textbf{Stationary state distribution}}                                                                                    & $E_i$ & $-\beta_{\substack{q_{\textrm{stat}}}}$ & $Z_{\substack{q_{\textrm{stat}}}} p\de{E_i}$                                                            \hfill (typically $q_{\substack{\textrm{stat}}} \geq 1$)     \\\hline
  \Espaco \raisebox{1.5ex}{\footnotesize{\textbf{Stationary state}}} \raisebox{-1.5ex}{\hspace{-2.56cm}\footnotesize{\textbf{distribution}}}         & $E_i$ & $-\beta_{\substack{q_{\textrm{stat}}}}$ & $Z_{\substack{q_{\textrm{stat}}}} p\de{E_i} = \el_{\substack{q_{\textrm{stat}}}}^{\displaystyle{-\beta_{\substack{q_{\textrm{stat}}}} E_i}}$                                                           \hfill (typically $q_{\substack{\textrm{stat}}} \geq 1$)   \\\hline
% \Espaco \footnotesize{\textbf{Sensitivity to initial conditions}}                                                                                & $t$   & $\lambda_{\substack{q_{\textrm{sen}}}}$ & $\xi = \el_{\substack{q_{\textrm{sen}}}}^{\displaystyle{\lambda_{\substack{q_{\textrm{sen}}}}\, t}}$    \hfill (typically $q_{\substack{\textrm{sen}}}  \leq 1$)     \\\hline
  \Espaco \raisebox{1.5ex}{\footnotesize{\textbf{Sensitivity to initial}}} \raisebox{-1.5ex}{\hspace{-3.11cm}\footnotesize{\textbf{conditions}}}   & $t$   & $\lambda_{\substack{q_{\textrm{sen}}}}$ & $\xi\de{t} = \el_{\substack{q_{\textrm{sen}}}}^{\displaystyle{\lambda_{\substack{q_{\textrm{sen}}}}\, t}}$   \hfill (typically $q_{\substack{\textrm{sen}}}  \leq 1$)   \\\hline
% \Espaco \footnotesize{\textbf{Typical relaxation of observable $O$} }                                                                            & $t$   & $-1/\tau_{\substack{q_{\textrm{rel}}}}$ & $\Omega = \el_{\substack{q_{\textrm{rel}}}}^{\displaystyle{-t/\tau_{\substack{q_{\textrm{rel}}}}\, t}}$ \hfill (typically $q_{\substack{\textrm{rel}}}  \geq 1$) \\\hline
  \Espaco \raisebox{1.5ex}{\footnotesize{\textbf{Typical relaxation}}} \raisebox{-1.5ex}{\hspace{-2.81cm}\footnotesize{\textbf{of observable $O$}}} & $t$   & $-1/\tau_{\substack{q_{\textrm{rel}}}}$ & $\Omega\de{t} = \el_{\substack{q_{\textrm{rel}}}}^{\displaystyle{-t/\tau_{\substack{q_{\textrm{rel}}}}\, t}}$ \hfill (typically $q_{\substack{\textrm{rel}}}  \geq 1$) \\\hline
%-----------------------------------------------------------------------
\end{tabular} 
\vspace{0.7cm}
\caption{{\it Top}: BG statistical mechanics, related to the differential equation (\ref{BGdiff}). {\it Bottom}: Nonextensive statistical mechanics, related to the differential equation (\ref{qdiff}). Typically we verify $q_{sen} \le 1 \le q_{stat} \le q_{rel}$. See details in \cite{Tsallis2004}. 
}
%\label{sample-figure}
%\end{minipage}
\label{qtriplet}
\end{table}

%\begin{figure}
%\begin{center}
%\resizebox*{14cm}{!}{\includegraphics{qtriplet.eps}}
%\end{center}
%\caption{{\it Top}: BG statistical mechanics. {\it Bottom}: Nonextensive statistical mechanics. Typically we verify $q_{sen} \le 1 \le q_{stat} \le q_{rel}$. See details in \cite{Tsallis2004}. 
%}
%\label{qtriplet}
%\end{figure}

The first verification of the existence of the $q$-triplet in nature came from NASA Goddard Space Flight Center \cite{BurlagaVinas2005} by using data of the spacecraft Voyager 1 on the solar wind.  Many more have since then been shown in very wide classes of natural and artificial systems.
The $q$-triplet in \cite{BurlagaVinas2005} is given by $(q_{sen},q_{stat},q_{rel})=(-0.6 \pm 0.2,1.75 \pm 0.06,3.8 \pm 0.3)$. These values have been dealt with $(q_{sen},q_{stat},q_{rel})=(-1/2,7/4,4)$  in \cite{TsallisGellMannSato2005b} as central elements of an infinite algebra constructed by the {\it additive duality} $q \to (2-q)$ and the {\it multiplicative duality} $q\to 1/q$; the case $q=1$ represents the fixed point of both dualities. By defining $\epsilon \equiv 1-q$, we obtain $(\epsilon_{sen},\epsilon_{stat},\epsilon_{rel})=(3/2,-3/4,-3)$ \cite{Baella2008}. These values satisfy amazing relations \cite{Baella2008}, namely
\begin{equation}
\epsilon_{stat}= \frac{\epsilon_{sen}+\epsilon_{rel}}{2} \;\;\; \textrm{(arithmetic mean)} \,,
\end{equation}
\begin{equation}
\epsilon_{sen}= \sqrt{\epsilon_{stat}\,\epsilon_{rel}} \;\;\;\textrm{(geometric mean)} \,,
\end{equation}
and
\begin{equation}
\frac{1}{\epsilon_{rel}}= \frac{1}{\epsilon_{sen}} +\frac{1}{\epsilon_{stat}} \;\;\;\textrm{(harmonic mean)} \,,
\end{equation}
the three Greek classical means! This fact remains up to now without any physical interpretation. These relations are {\it not} verified in other $q$-triplets available in the literature. Therefore they seem to characterize some unknown special conditions, perhaps related to some symmetry properties.

Another interesting $q$-triplet also is available, namely for the logistic map at its edge of chaos: $(q_{sen},q_{stat},q_{rel})=(0.24448...,1.65 \pm 0.05,2.249784109...)$. Once again, Baella heuristically found a remarkable relation \cite{Baella2010}:
\begin{equation}
\epsilon_{sen} + \epsilon_{rel}=\epsilon_{sen}+\epsilon_{stat} \,.
\label{baella2010}
\end{equation}
This relation implies $q_{stat}=\frac{\epsilon_{del}-1}{1-\epsilon_{sen}}=1.65424$, which compares remarkably well with $1.65 \pm 0.05$. Once again, relation (\ref{baella2010}) eludes any physical/mathematical interpretation up to now.

These examples illustrate that, for a given thermostatistical system, there exist not one but an infinite set of $q$-indices, each of them characterizing a certain class of properties, the most relevant among them being those appearing in the $q$-triplet. Only very few of those indices, say a couple of them, are independent, all the others most probably being functions of those few. These general functions constitute nowadays an important open point. Several important hints are nevertheless available in the literature, whose detailed description escapes to the scope of the present overview.

\section{Applications}

A vast variety of predictions, verifications, applications are available in the literature for natural, artificial and social systems, through analytical, experimental, observational and computational efforts. A regularly updated Bibliography can be accessed at \cite{biblio}. Here we will restrict ourselves to a few selected such applications. Many more can however be seen in \cite{AnteneodoTsallis1997,TamaritCannasTsallis1998,NobreTsallis1995,TsallisLenzi2002,TsallisAnjosBorges2003,CombeRichefeuViggianiHallTengattiniAtman2013,RuizBountisTsallis2012,GougamTribeche2011,RiosGalvaoCirto2011,GougamTribeche2011b,DouglasBergaminiRenzoni2006,LiuGoree2008,PickupCywinskiPappasFaragoFouquet2009,DeVoe2009,BurlagaVinasNessAcuna2006,BurlagaNess2011,EsquivelLazarian2010,LivadiotisMcComas2011,NobreMonteiroTsallis2011,NobreMonteiroTsallis2012,SoaresTsallisMarizSilva2005,ThurnerTsallis2005,WhiteKejzarTsallisFarmerWhite2006,ThurnerKyriakopoulosTsallis2007,Hasegawa2006}. \\

\begin{figure}
\begin{center}
%\begin{minipage}{100mm}
%\subfigure[]{
\resizebox*{11cm}{!}{\includegraphics{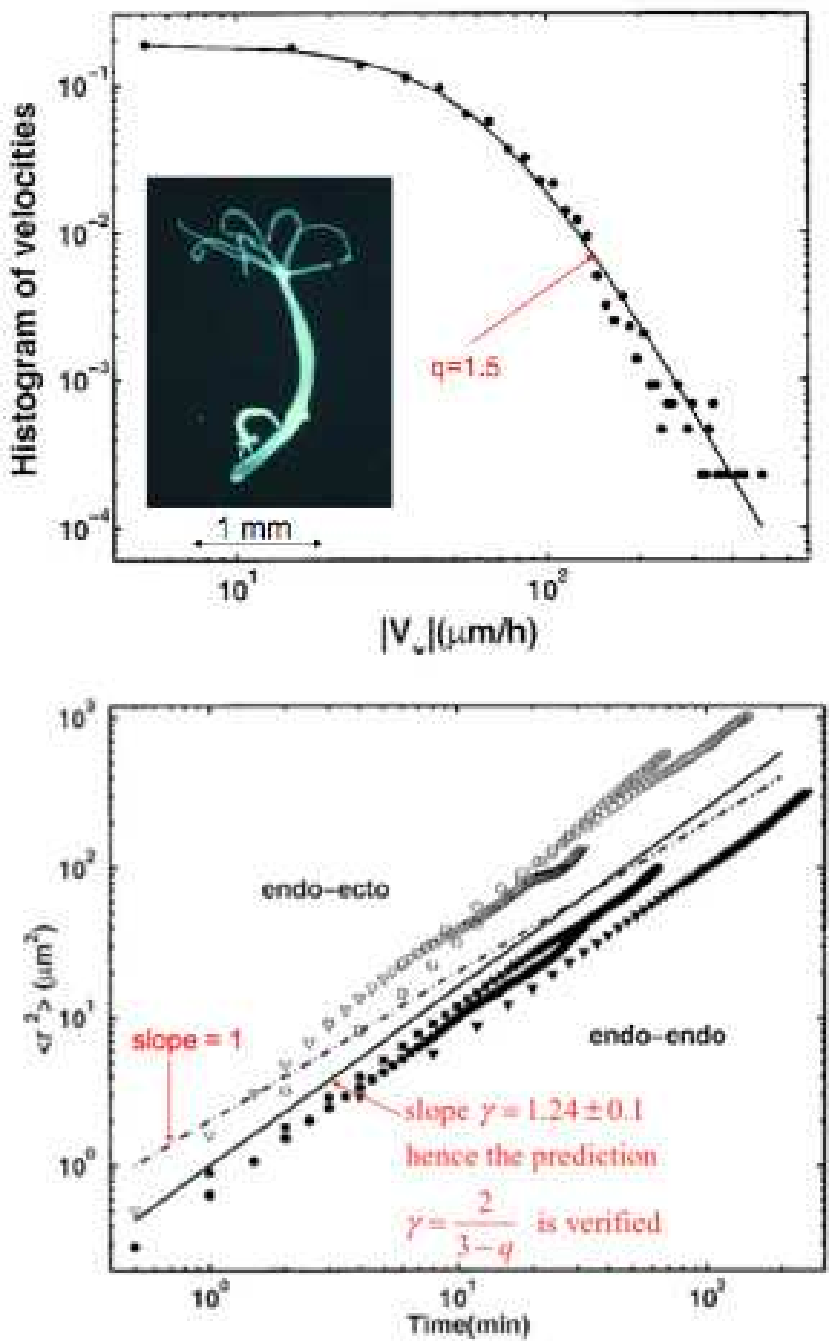}}
%\subfigure[]{
%\resizebox*{5cm}{!}{\includegraphics{senu_gr2.eps}}}%
\end{center}
\caption{{\it Top}: $q$-Gaussian distribution of velocities of {\it Hydra viridissima} cells. {\it Bottom}: Anomalous diffusion of the same cells. The scaling prediction $\gamma = 2/(3-q)$ of \cite{TsallisBukman1996} is verified. See details in \cite{UpadhyayaRieuGlazierSawada2001}.
}
%\label{sample-figure}
%\end{minipage}
\label{hydra}
\end{figure}

(i) The first experimental verification of $q$-statistics concerned a living organism, namely {\it Hydra viridissima} \cite{UpadhyayaRieuGlazierSawada2001}: the velocity distribution was found to be non-Maxwellian, more precisely a $q$-Gaussian with $q \simeq 1.5$. Moreover, anomalous diffusion was also measured with $x^2$ scaling like $t^\gamma$ with $\gamma$ satisfying, within error bars, the predicted scaling law $\gamma = 2/(3-q)$ \cite{TsallisBukman1996}. See Fig. \ref{hydra}. \\

\begin{figure}
\begin{center}
%\begin{minipage}{100mm}
%\subfigure[]{
\resizebox*{12cm}{!}{\includegraphics{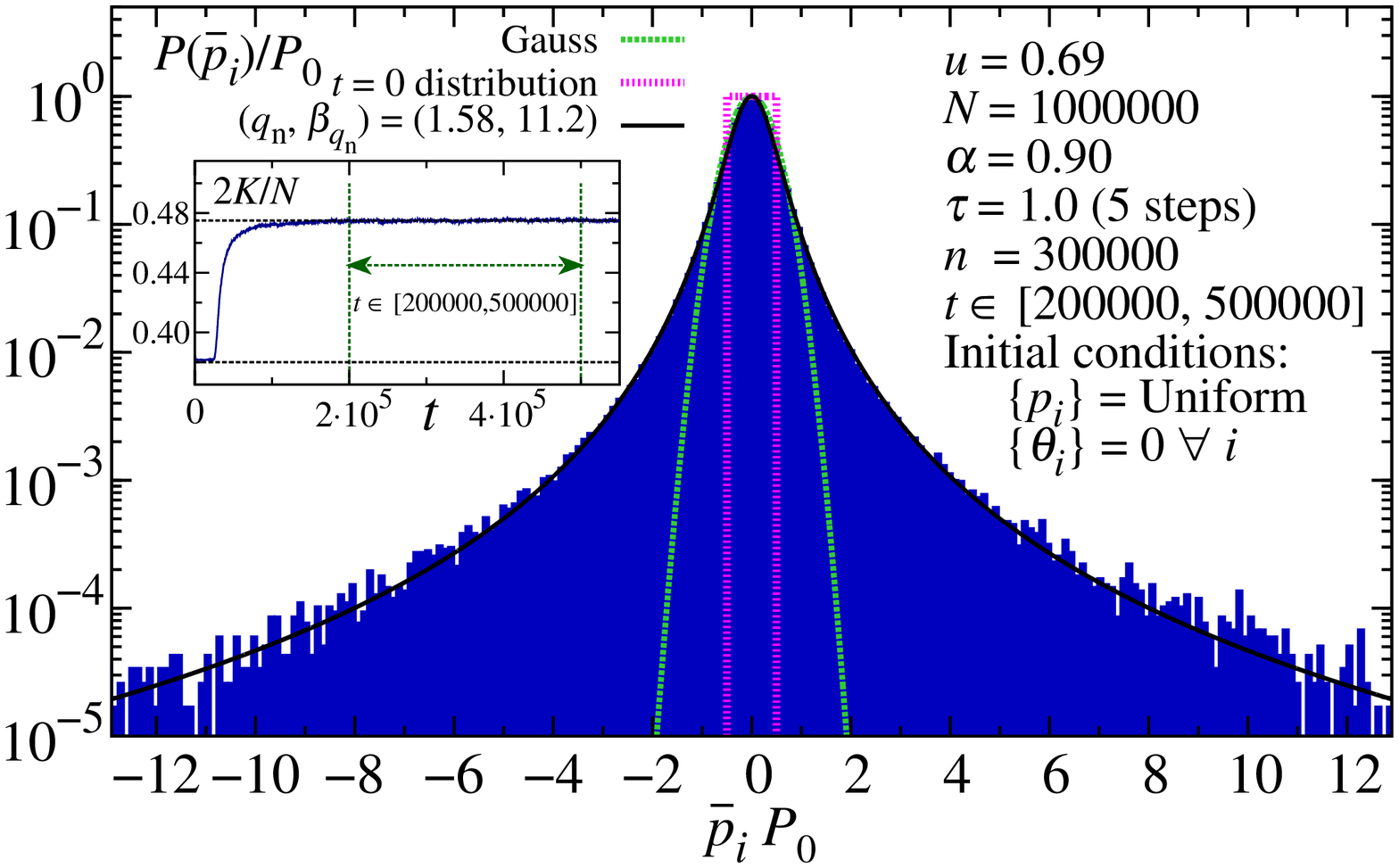}}
\resizebox*{12cm}{!}{\includegraphics{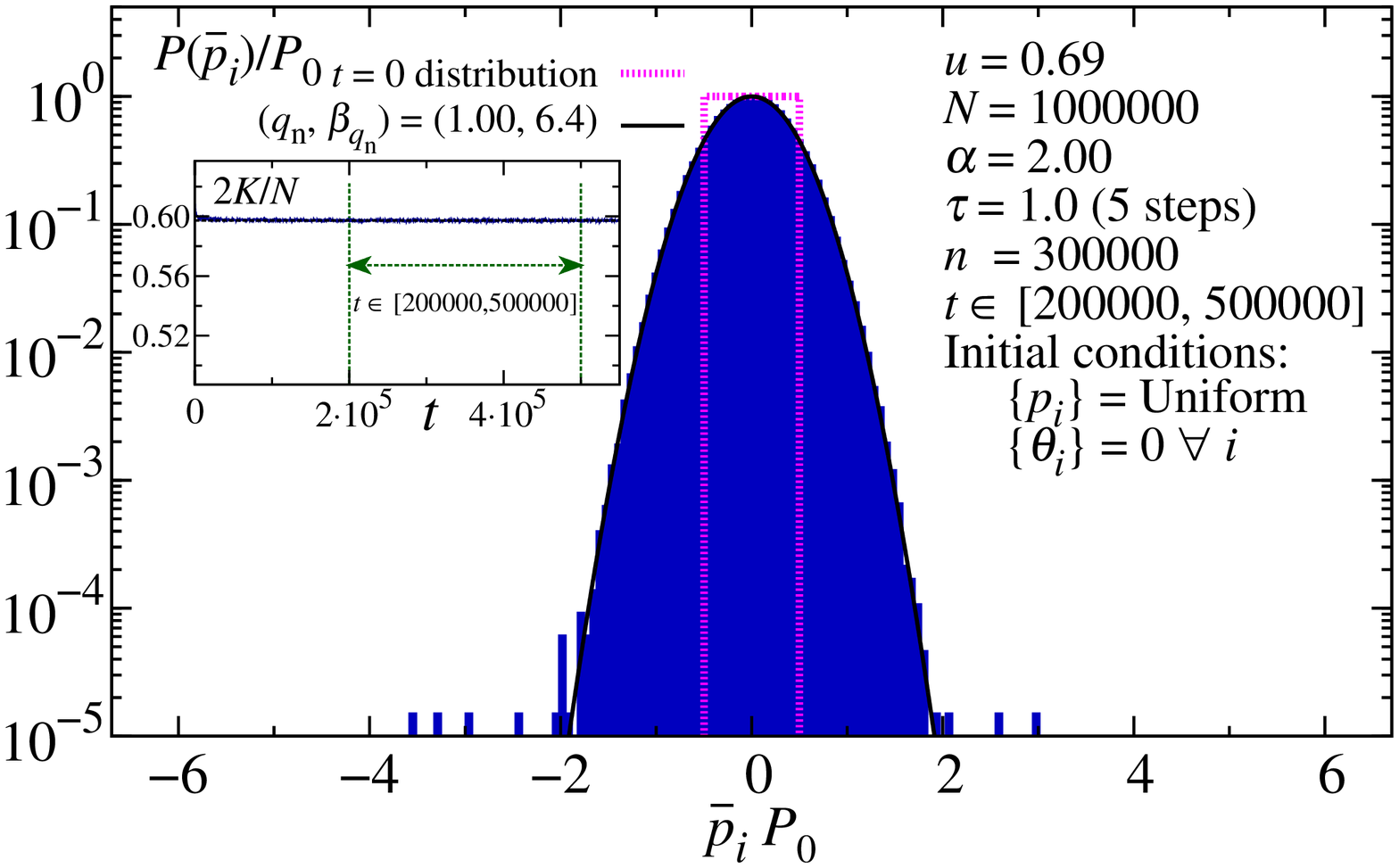}}
%\subfigure[]{
%\resizebox*{5cm}{!}{\includegraphics{senu_gr2.eps}}}%
\end{center}
\caption{Time-averaged velocity distribution for large values of time. {\it Top}: $\alpha =0.9$ (long-range interactions); {\it Bottom}:  $\alpha=2$ (short-range interactions). See details in \cite{CirtoAssisTsallis2013}.
}
%\label{sample-figure}
%\end{minipage}
\label{qgaussian}
\end{figure}

\begin{figure}[t]
\begin{center}
%\begin{minipage}{100mm}
%\subfigure[]{
\resizebox*{11cm}{!}{\includegraphics{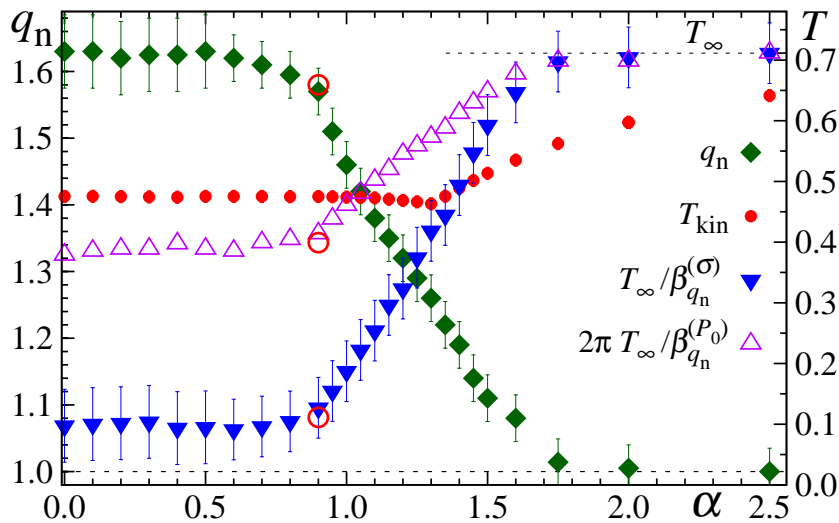}}
%\subfigure[]{
%\resizebox*{5cm}{!}{\includegraphics{senu_gr2.eps}}}%
\end{center}
\caption{The $q$-index exhibiting that, for $\alpha$ small enough (long-range interactions), the time-averaged velocity distribution is, for large values of time, quite well approached by $q$-Gaussians with $q>1$, whereas, for $\alpha$ large enough (short-range interactions), the distribution is Maxwellian. The fact that the crossing does not occur precisely at $\alpha=1$ is believed to be a finite-size effect. The red dots correspond to the {\it kinetic temperature} (proportional to the variance of the velocities). The two sets of triangles correspond to a (unique) different temperature (sometimes referred to as the {\it effective temperature}, and characterizing the width of a non-Gaussian distribution), normalized in two different manners. In the thermodynamic limit ($N \to\infty$), one expects the kinetic and the effective temperatures to coincide for $\alpha >1$, and to be different for $0 \le \alpha <1$. However, at the present stage, this remains as an open point. See details in \cite{CirtoAssisTsallis2013}.
}
%\label{sample-figure}
%\end{minipage}
\label{qalphaplot}
\end{figure}

(ii) A recent numerical illustration of how ergodicity and nonergodicity are crucially relevant to the validity or violation of BG statistical mechanics is available. It concerns the effects of long-range interactions in classical many-body Hamiltonian systems \cite{PluchinoRapisardaTsallis2007,CirtoAssisTsallis2013}. The system which is focused is a one-dimensional chain (i.e., $d=1$) of $N$ planar rotators interacting through attractive interactions decaying as $1/r^\alpha \; (\alpha \ge 0)$. The potential is integrable for $\alpha>1$, and nonintegrable otherwise. It has long been shown (numerically) that,  in the $N\to\infty$ limit, the maximal Lyapunov exponent is positive for $\alpha>1$ and vanishes for $0 \le \alpha \le 1$ \cite{AnteneodoTsallis1998}. At very large times, the ensemble-averaged velocity distribution is Gaussian for all values of $\alpha$. In remarkable contrast, the time-averaged velocity distribution is Maxwellian ($q$-Gaussian with $q>1$) for $\alpha$ large (small) enough: see Figs. \ref{qgaussian} and \ref{qalphaplot}. \\

\begin{figure}
\begin{center}
%\begin{minipage}{100mm}
%\subfigure[]{
\resizebox*{11cm}{!}{\includegraphics{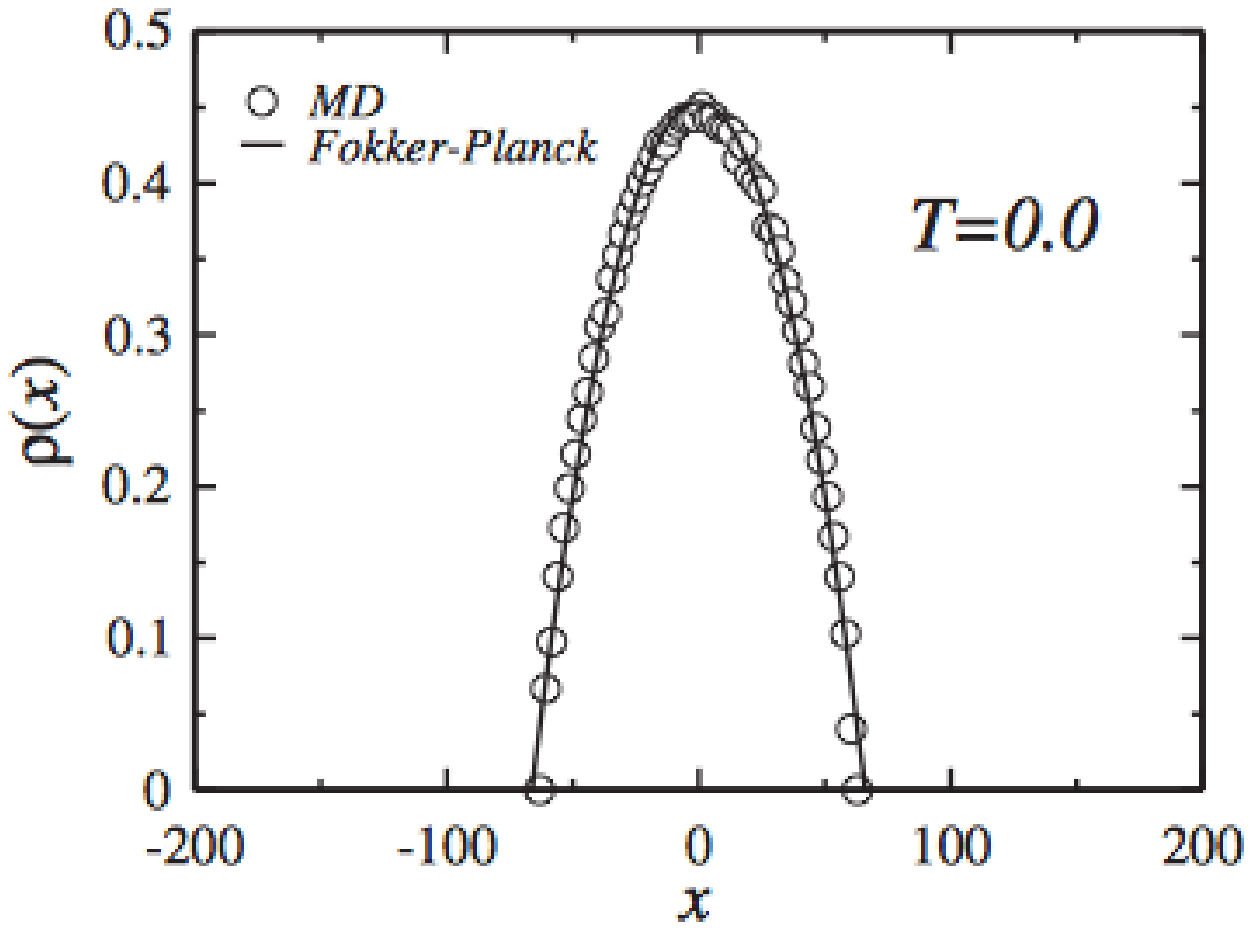}}
\resizebox*{11cm}{!}{\includegraphics{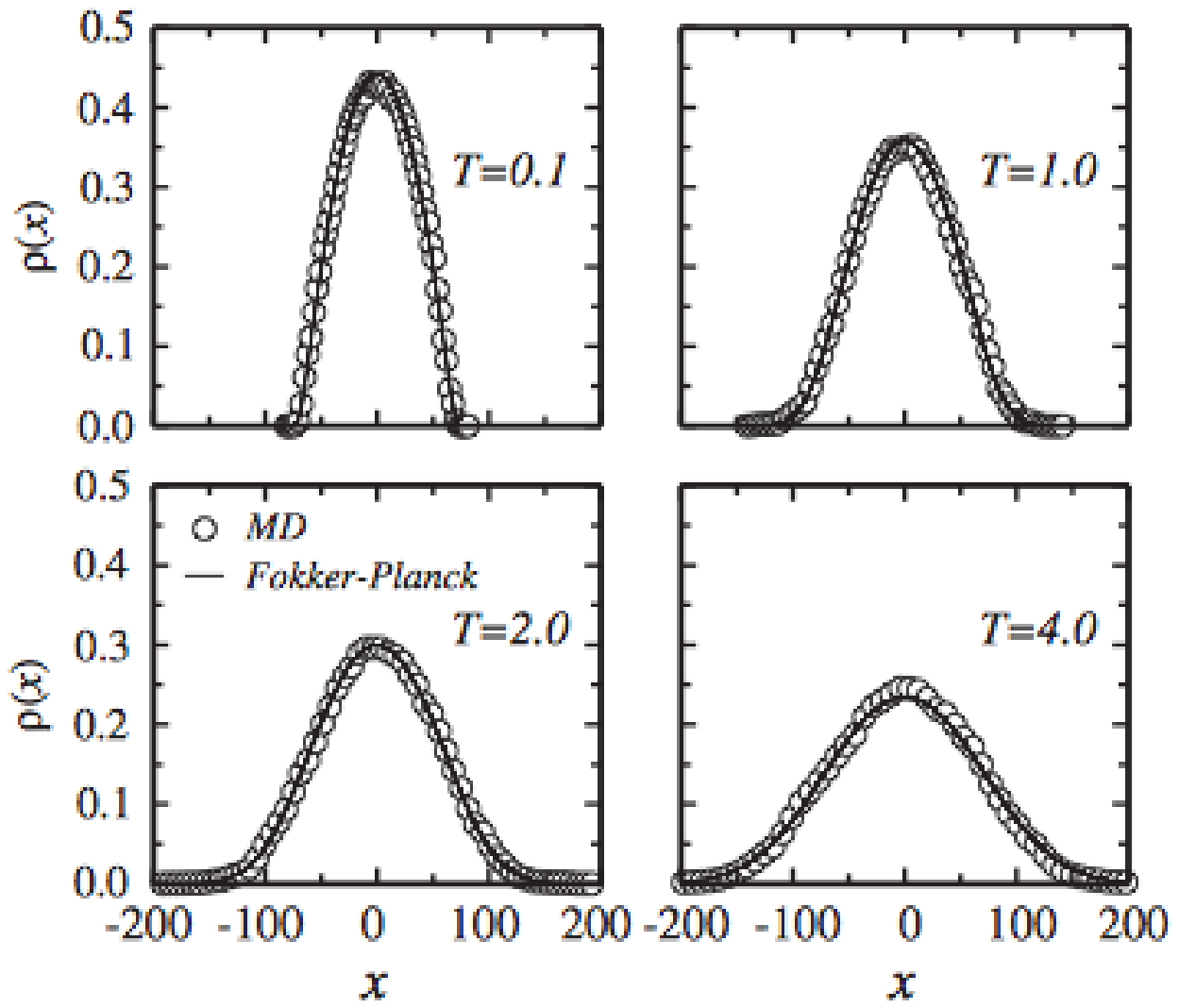}}
%\subfigure[]{
%\resizebox*{5cm}{!}{\includegraphics{senu_gr2.eps}}}%
\end{center}
\caption{The stationary-state space distribution for kinetic temperature $T=0$ in the presence of an external confining harmonic potential is analytically proved to be a $q$-Gaussian with $q=0$. For increasing values of $T$ the distribution gradually approaches a Gaussian form, which precisely corresponds to the $T\to\infty$ limit. In all cases, the molecular-dynamical data confirm the analytical results. From \cite{AndradeSilvaMoreiraNobreCurado2010}.  
}
%\label{sample-figure}
%\end{minipage}
\label{PRL}
\end{figure}

\begin{figure}
\begin{center}
%\begin{minipage}{100mm}
%\subfigure[]{
\resizebox*{11cm}{!}{\includegraphics{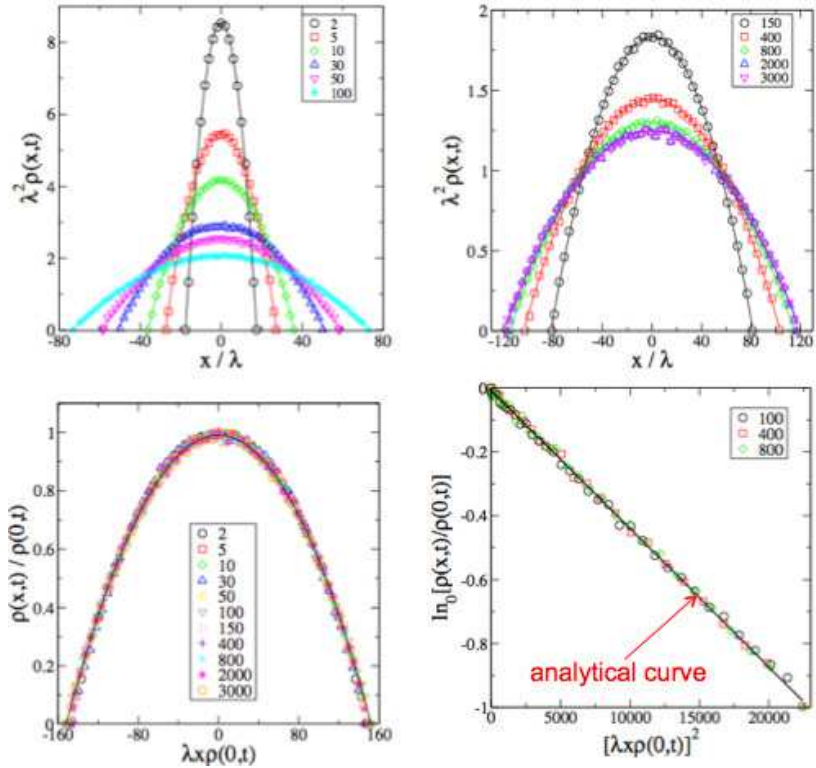}}
%\subfigure[]{
%\resizebox*{5cm}{!}{\includegraphics{senu_gr2.eps}}}%
\end{center}
\caption{The space distributions for typical values of times $t$ for kinetic temperature $T=0$ in the presence of an external confining harmonic potential are analytically proved to be $q$-Gaussians with $q=0$. In all cases, the molecular-dynamical data confirm the analytical results. One of the panels exhibits perfect data collapse in linear-linear representation as well as in $(q=0)$-logarithmic - quadratic representation. From \cite{RibeiroNobreCurado2012b}.  
}
%\label{sample-figure}
%\end{minipage}
\label{PREx}
\end{figure}

\begin{figure}[t]
\begin{center}
%\begin{minipage}{100mm}
%\subfigure[]{
\resizebox*{11cm}{!}{\includegraphics{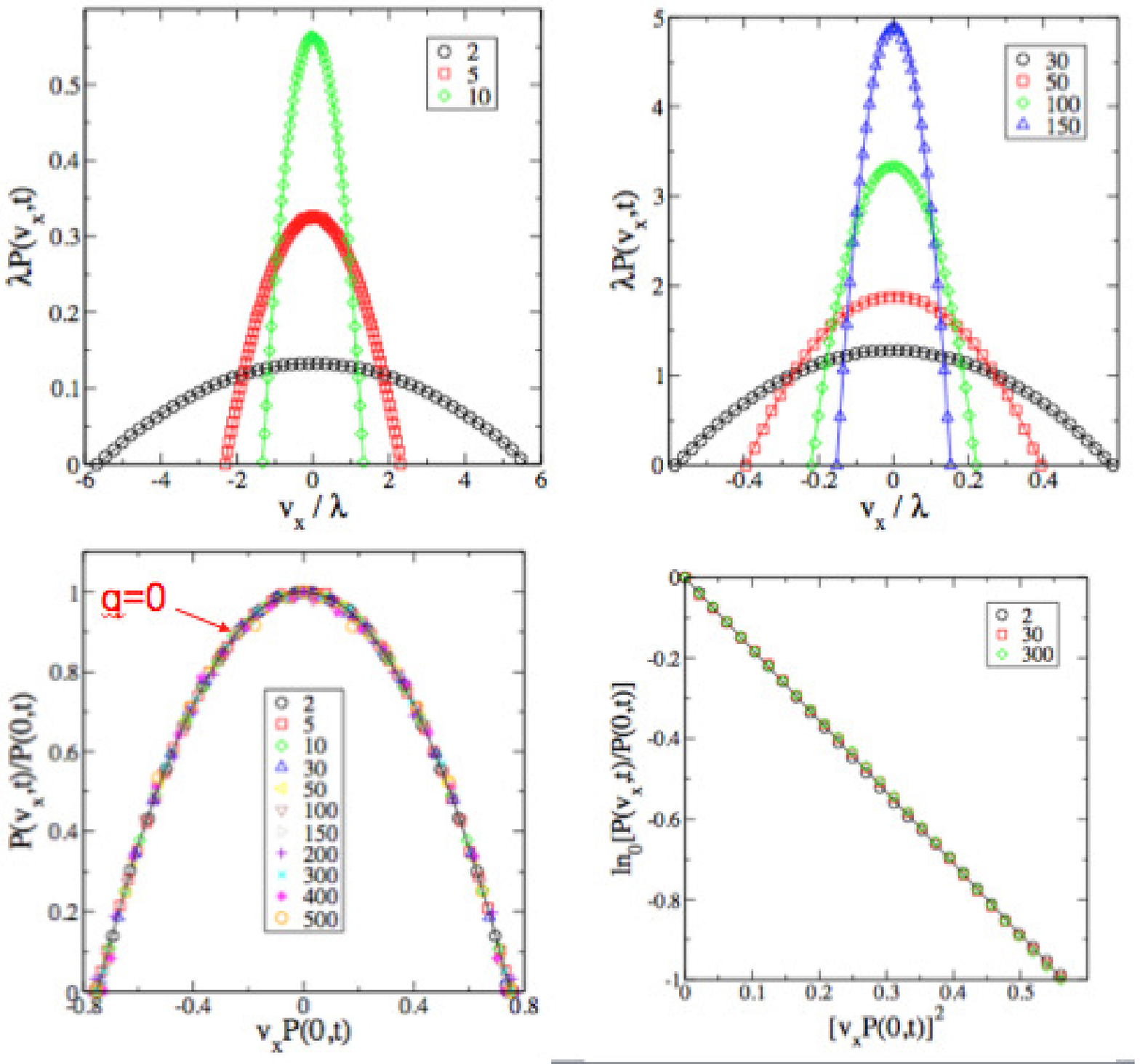}}
%\subfigure[]{
%\resizebox*{5cm}{!}{\includegraphics{senu_gr2.eps}}}%
\end{center}
\caption{The velocity distributions for typical values of times $t$ for kinetic temperature $T=0$ through molecular-dynamics. One of the panels exhibits visible data collapse in linear-linear representation as well as in $(q=0)$-logarithmic - quadratic representation. This results neatly exhibits a non-Boltzmannian behavior which, for all Hamiltonian systems, yields a Gaussian form ($q=1$). From \cite{RibeiroNobreCurado2012b}.  
}
%\label{sample-figure}
%\end{minipage}
\label{PREv}
\end{figure}

(iii) Another interesting recent application concerns a many-body dissipative system which mimics the {\it overdamped} motion of (repulsively) interacting vortices in type-II superconductors \cite{AndradeSilvaMoreiraNobreCurado2010,RibeiroNobreCurado2011,RibeiroNobreCurado2012,RibeiroNobreCurado2012b,NobreSouzaCurado2012,CasasNobreCurado2012}. Both the distribution of positions and that of velocities are described, at zero {\it kinetic temperature} $T$ (though nonzero {\it effective temperature} $\theta$), by $q$-Gaussians with $q=0$, in strong contradiction with the Maxwellian distribution of velocities ($q=1$) expected within BG statistical mechanics. See  the respective anomalous distributions in Figs. \ref{PRL}, \ref{PREx} and \ref{PREv}. Furthermore, the validity of the Carnot cycle has been recently established in terms of the effective temperatures  in such systems \cite{CuradoSouzaNobreAndrade2014}. The efficiency is proved to be $\eta = 1-\frac{\theta_2}{\theta_1}$ \cite{CuradoSouzaNobreAndrade2014}, where $\theta_1$ and $\theta_2$ represent the effective temperatures associated to the ÒhotterÓ (higher vortex density) and ÒcolderÓ (lower vortex density) heat reservoirs.\\

\begin{figure}
\begin{center}
%\begin{minipage}{100mm}
%\subfigure[]{
\resizebox*{11cm}{!}{\includegraphics{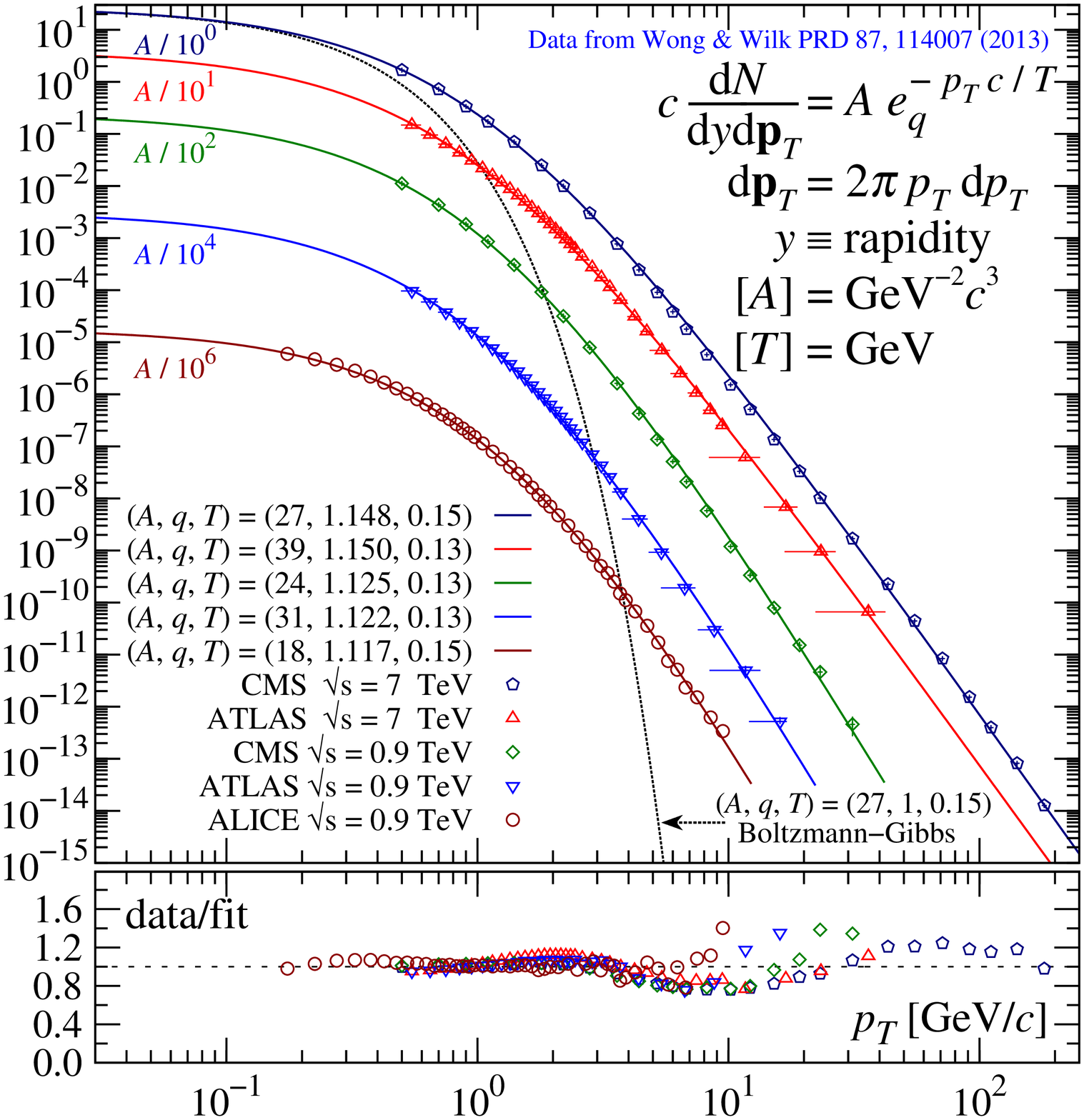}}
%\subfigure[]{
%\resizebox*{5cm}{!}{\includegraphics{senu_gr2.eps}}}%
\end{center}
\caption{The distribution of hadronic transverse momenta in various experiments of proton-proton collisions. The data are taken from \cite{WongWilk2013}. The continuous curves are simple $q$-exponentials. From the region with probability close to one on, the present $q \simeq 1.1$ curves depart from the BG ($q=1$) ones. If we start counting from that point on, the remarkable agreement continues along 14 decades. From \cite{CirtoTsallis2013}. 
}
%\label{sample-figure}
%\end{minipage}
\label{highenergy}
\end{figure}

\begin{figure}
\begin{center}
%\begin{minipage}{100mm}
%\subfigure[]{
\resizebox*{12cm}{!}{\includegraphics{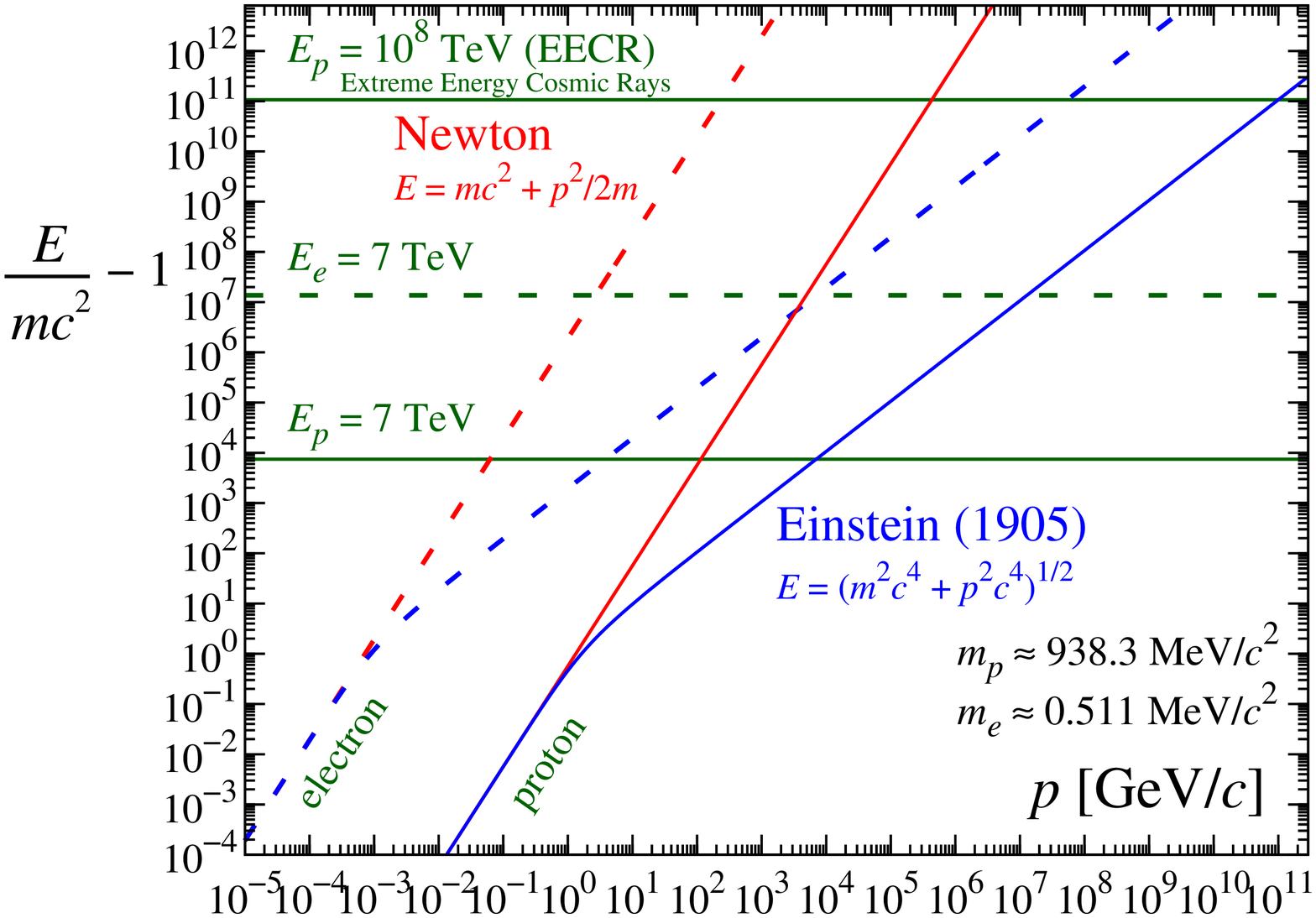}}
%\subfigure[]{
%\resizebox*{5cm}{!}{\includegraphics{senu_gr2.eps}}}%
\end{center}
\caption{The Newtonian and relativistic expressions for the kinetic energy $E$ of a free particle of mass $m$ as functions of the momentum $p$. From the region with $E_p/mc^2 -1$ close to one on, the Einstein curves depart from the Newtonian ones. If we start counting from that point on up to the energies of the most energetic observed cosmic rays, the relativistic curves have been experimentally verified along 11 decades.}
%\label{sample-figure}
%\end{minipage}
\label{einstein}
\end{figure}

(iv) A considerable amount of papers have explored the fact that high-energy collisions of elementary particles or heavy ions at LHC/CERN (CMS, ALICE, ATLAS detectors) and RHIC/Brookhaven (STAR, PHENIX detectors) yield results that are quantitatively consistent with $q$-statistics \cite{BediagaCuradoMiranda2000,Beck2000,Khachatryan2010,
%Khachatryan2010b,Khachatryan2010c,Khachatryan2010d,
Chatrchyan2011,Aamodt2010,
%Aamodt2011,Aamodt2011b,
Abelev2012,Aad2011,Adare2011,
%Adare2011b,
%Adare2011c
Adare2011d,ShaoYiTangChenLiXu2010,WongWilk2012,Deppman2012,MarquesAndradeDeppman2013}. This is particularly so for the transverse momenta distributions of hadronic jets resulting from proton-proton collisions: See Fig. \ref{highenergy}. \\
As we see the quantitative agreement holds along amazing 14 decades for the probabilities! In order to make some sort of comparison, and given the fact that very few phenomena in nature can be observed along so many decades, we have checked how many experimental decades are available nowadays for the Einstein expression for the kinetic energy $E=\sqrt{m^2 c^4 + p^2 c^2}$, departing from Newton expression $E=p^2/2m$: see Fig. \ref{einstein}. \\

\begin{figure}
\begin{center}
%\begin{minipage}{100mm}
%\subfigure[]{
\resizebox*{11cm}{!}{\includegraphics{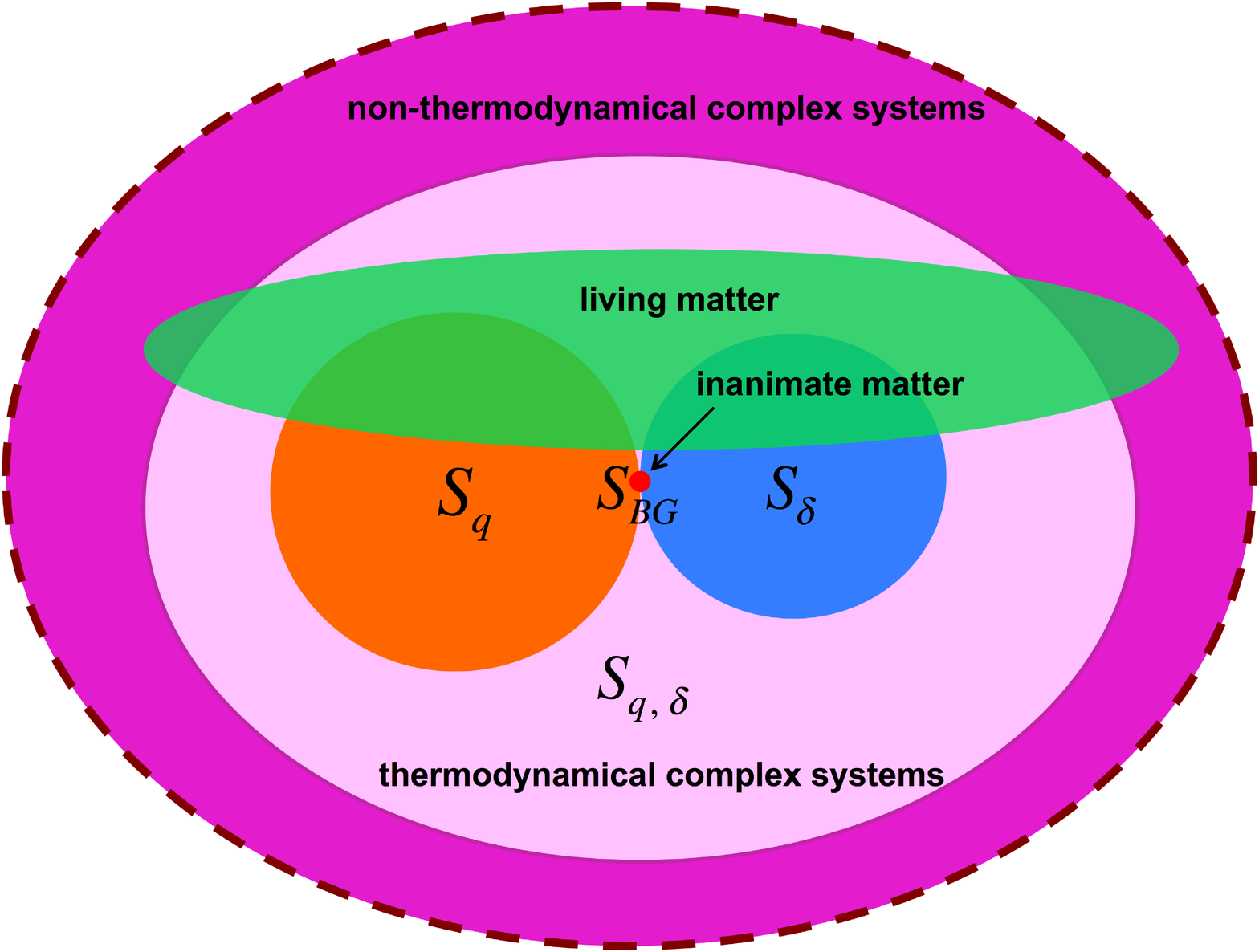}}
%\subfigure[]{
%\resizebox*{5cm}{!}{\includegraphics{senu_gr2.eps}}}%
\end{center}
\caption{A thinking scenario to represent {\it inanimate} versus {\it living} matter. The entropy $S_{q,\delta}$ recovers $S_q$ for $\delta=1$, $S_\delta$ for $q=1$, and $S_{BG}$ for $(q,\delta)=(1,1)$. The values of the indices $(q,\delta)$ are chosen in such a way that thermodynamics is satisfied, i.e., $S_{q,\delta}(N) \propto N$, or equivalently $S_{q,\delta}(L) \propto L^d$, where $d$ is the integer or fractal dimension of the system (with $N \propto L^d$), $L$ being a linear size of the system. $S_{q,\delta}$ appears to be thermodynamically equivalent to $S_{c,d}$ given by Eq. (\ref{cdentropy}). Inanimate matter --- basically {\it ergodic} --- is primarily linked to the additive entropy $S_{BG}$. In contrast, living matter --- basically {\it nonergodic} --- appears to be primarily linked to nonadditive entropies ($S_q,\,S_{\delta}, \, S_{q,\delta}$ with $(q,\delta) \ne (1,1)$) whenever a thermodynamical approach is appropriate for a specific aspect, but it is kind of natural to imagine that, for other aspects, a non-themodynamical realm might be necessary.  In such a (considerable simplified) view, inanimate matter strictly refers say to a crystal, or to a simple fluid, not to a glass or amorphous or granular matter, whose very slow and intricate evolution makes them to be somehow better described as a complex, living-like system. Analogously, the evolution of a language or of an economical system surely has many common aspects with living matter. Developments in non-equilibrium thermodynamics from a different standpoint have been reviewed in \cite{JouRestuccia2011}.
}
%\label{sample-figure}
%\end{minipage}
\label{livingversusinanimate}
\end{figure}

(v) A typical (3+1)-dimensional black-hole would in principle be expected to have an entropy proportional to $L^3$, $L$ being its characteristic linear size. There is however a vast literature \cite{Bekenstein1973,
%Bekenstein1973_2,
Hawking1974,Hawking1974_2,tHooft1985,
%tHooft1990,
Susskind1993,Maddox1993,Srednicki1993,
%StromingerVafa1996,
MaldacenaStrominger1998,DasandShankaranarayanan2006,BrusteinEinhornYarom2006,
%BorstenDahanayakeDuffEbrahimRubens2009,
Padmanabhan2009,
%Casini2009,
BorstenDahanayakeDuffMarraniRubens2010,Corda2011,KolekarPadmanabhan2011,Saida2011} stating that the black-hole entropy is (intriguingly) proportional to $L^2$. There are indeed many (quite convincing) physical arguments, for instance those related to the holographic principle, that lead to $S_{BG} \propto L^2$.  To solve this puzzle, it has been recently advanced \cite{TsallisCirto2013} that this non-thermodynamic behavior can be overcome if we adopt, as thermodynamic entropy, not the BG functional, but the $S_\delta$ one for $\delta=3/2$. In this case, we recover standard entropic extensivity, since we verify that $S_{\delta=3/2}(L) \propto L^3$. In a recent paper \cite{KomatsuKimura2013} it has been shown various physical consequences associated with this nonadditive entropy. One of them is that, in contrast with $S_{BG}$, $S_\delta$ appears to imply that the concept of "dark energy" might be not necessary in order to explain the present cosmological observations of the accelerating expansion of the universe.

%(vi) Several methods to process images, signals and other data have sensibly benefited from the use of $S_q$ and related concepts. This is so in particular for medical procedures: see for instance  \cite{DinizMurtaBrumAraujoSantos2010,ShiLiMiaoHu2012,CapurroDiambraLorenzoMacadarMartinMostaccioPlastinoRofmanTorresVelluti1998,CapurroDiambraLorenzoMacadarMartinsMostaccioPlastinoPerezRofmanTorresVelluti1999,TorresRufinerMiloneCherniz2007,LopesOliveiraCesar2009,ZhangWu2008,SotolongoGrauRodriguezPerezAntoranzSotolongoCosta2010,SotolongoGrauRodriguezPerezSotolongoCostaAntoranz2013}. An interesting such example concerns a technique for the automatic detection of microcalcifications from mammograms \cite{MohanalinBeenamolKalraKumar2010}, as shown  in Fig. \ref{image}.

\section{Concluding remarks on how entropy reflects the inanimate or living nature of matter}

We may summarize the present overview as follows. The systems which we may strictly consider as inanimate matter (e.g., a crystal, a simple fluid) typically visit, along time, nearly all the states of the appropriate phase space (or a {\it finite} fraction of those) with equal frequency, i.e., they are {\it ergodic} in a region of phase space with finite Lebesgue measure. As such, they relatively quickly approach their thermal equilibrium whenever isolated or in contact with a thermostat. Their thermal equilibrium is well described by the BG entropy and its associated statistical mechanics. Its macroscopic behaviour is nearly the same starting from any initial condition. i.e., the initial condition of the system is quickly forgotten.

Living (e.g., a bacteria, an animal) or living-like (e.g., a language, an urban organization) matter is definitively different. It is basically {\it nonergodic}, and either it is in a stationary (or quasi-stationary) state or it slowly approaches such a state (generically quite different from a thermal equilibrium). The initial condition is either not forgotten or forgotten extremely slowly, even if for a wide class of initial conditions the individual differences are relatively modest. The system lives in a subset of phase space with zero Lebesgue measure (e.g., a multifractal).  The corresponding stationary or quasi stationary states are frequently well described by nonadditive entropies such as the $q$-entropy and the $\delta$-entropy and their associated statistical mechanics. In spite of the individual differences being modest, they can be important at the level of the evolution. Indeed, there can be no evolution without diversity.

The reader may allow us to appeal to a metaphor --- ``of all things the greatest", according to Aristotle! ---. If we have say a ``super-fly" traveling around the world in Brownian motion, its trajectory will eventually cover the entire planet with nearly uniform probability. This distribution will be the same for different starting points of the flights. The Lebesgue measure will roughly be the area of the planet surface, conceived as an Euclidean quasi-sphere. This is the dynamical-geometrical scenario for Boltzmann-Gibbs entropy and statistical mechanics. If we consider instead the flights of Air France, they constitute an hierarchical structure that never forgets that it is centered in Paris, its most important hub. Analogously Iberia is centered in Madrid, and British Airways in London.  They are all different in what concerns their centers and the regions they cover, but they are all quite similar in what concerns their dynamics and geometry. Their Lebesgue measure is zero (if we consider the airports as points). This is the scenario for the nonadditive entropies $S_q$, $S_\delta$, $S_{q.\delta}$, and their corresponding statistical mechanics.

In other words, within the present statistical-mechanical context, the game ``inanimate {\it versus} living" is somehow the game ``thermal equilibrium {\it versus} slowly evolving quasi-stationary state". The first one surely is adequately described in terms of the BG entropy, whereas the second mandates nonadditive entropies like $S_q$ or $S_\delta$ or even $S_{q,\delta}$. If (and, paraphrasing Darwin \cite{Darwin1871}, oh what a big if) it would be possible to classify on entropic grounds very rich concepts such as complexity, or usefully distinguish inanimate from living matter, we might imagine, at least as a schematic scenario, something like Fig. \ref{livingversusinanimate}.

\subsection*{Acknowledgements}

R. Ferreira was the first to point out (in a telephone conversation we had long ago) that $S_q$ could somehow reflect, for $q \ne 1$, the special nature of living matter. T. Bountis, L.J.L. Cirto, E.M.F. Curado, R. Hanel, F.D. Nobre, M. Lopes da Rocha, R. Maynard, P. Tempesta, S. Thurner, G. Wilk and C.Y. Wong are warmly acknowledged for fruitful discussions. Stimulating comments by A. Stefanovska have warmly encouraged me. By courtesy of their authors, some of the figures reproduced here come from various articles, indicated case by case.  
I also acknowledge partial financial support by CNPq, FAPERJ and CAPES (Brazilian agencies), and by the European Union (European Social Fund Ð ESF) and Greek national funds through the Operational Program `Education and Lifelong Learning" of the National Strategic Reference Framework (NSRF) - Research Funding Program: {\it Thales. Investing in knowledge society through the European Social Fund}.

\clearpage

\subsection*{Notes on contributor}
Constantino Tsallis is a Professor of Physics, Head of the National Institute of Science and Technology for Complex Systems of Brazil, Emeritus Researcher at the Brazilian Center for Research in Physics, and External Faculty at the Santa Fe Institute, New Mexico.  He obtained in 1974 his Doctorat d'Etat \`es Sciences Physiques at the University of Paris-Orsay, and has been honored with Doctor Honoris Causa in several countries, among other Brazilian and international prizes and distinctions. He has published over 300 articles and books, and proposed in 1988 a generalisation of Boltzmann-Gibbs entropy and statistical mechanics.
\begin{figure}
\includegraphics[width=0.2\linewidth]{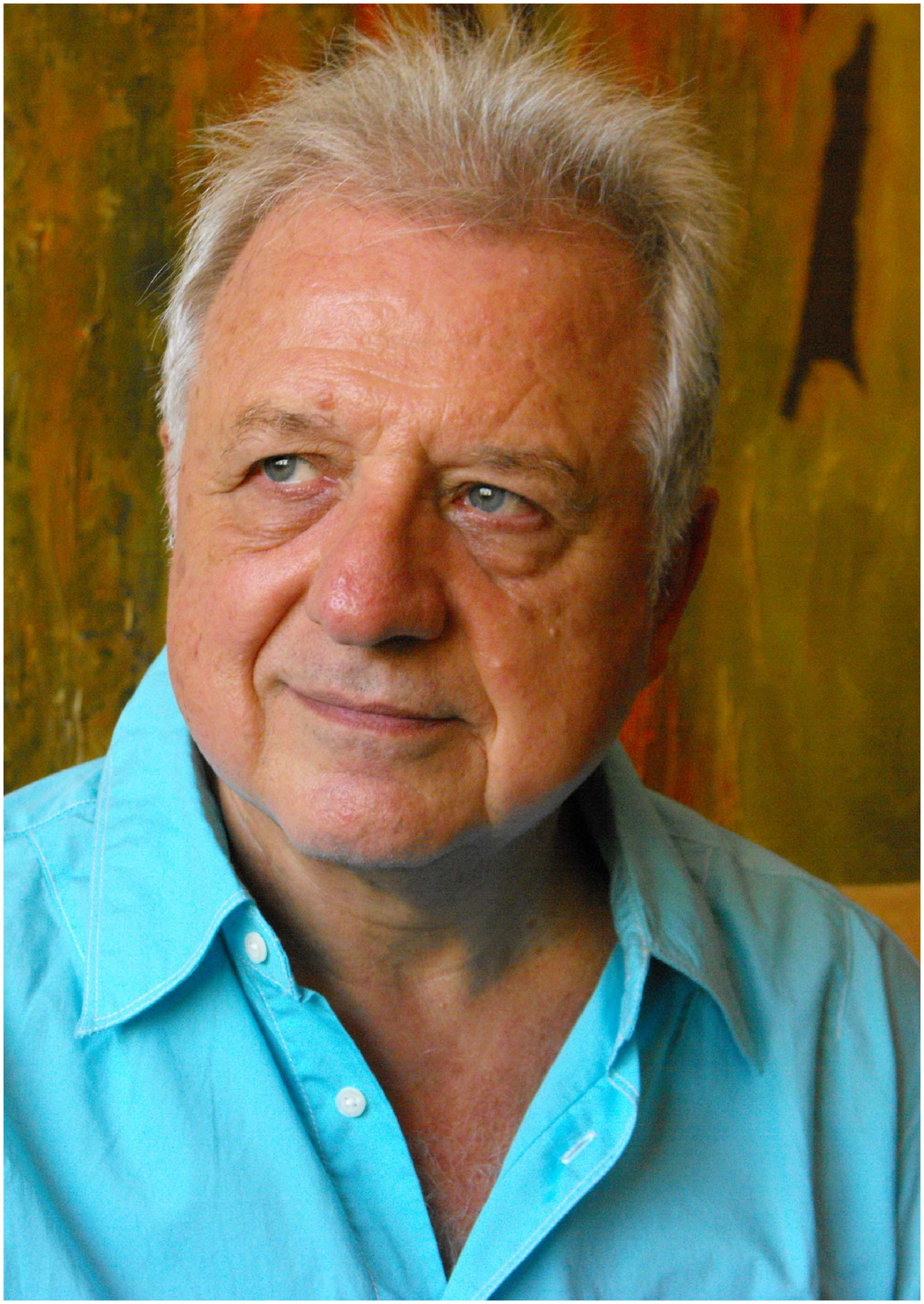}
\end{figure}

\medskip
\

\label{lastpage}

\end{document}